\documentclass[12pt,preprint]{aastex}

\shorttitle{Lengths, Colors and Ages of Bars}
\shortauthors{Gadotti \& de Souza}

\begin{document}

\title{On the Lengths, Colors and Ages of 18 Face--On Bars}
\author{D. A. Gadotti\altaffilmark{1,2,3}}
\email{dimitri@mpa-garching.mpg.de}
\and
\author{R. E. de Souza\altaffilmark{1}}
\email{ronaldo@astro.iag.usp.br}
\altaffiltext{1}{Departamento de Astronomia, Universidade de S\~ao Paulo, Rua do Mat\~ao 1226,
05508-090, S\~ao Paulo-SP, Brasil}
\altaffiltext{2}{Laboratoire d'Astrophysique de Marseille,
2 Place Le Verrier, 13248 Marseille Cedex 04, France}
\altaffiltext{3}{Present Address: Max-Planck-Institut f\"ur Astrophysik, Karl-Schwarzschild-Str. 1, D-85748
Garching bei M\"unchen, Germany}

\begin{abstract}
Along with a brief analysis we present data obtained from BVRI and Ks images of a sample of 19 galaxies
(18 barred and 1 unbarred) which will be further explored in a future paper. We measured
the lengths and colors of the bars, created color maps and estimated global color gradients.
Applying a method developed in a companion paper, we could distinguish for 7 galaxies in
our sample those whose bars have been recently formed from the ones with already evolved bars. We estimated
an average difference in the optical colors between young and evolved bars that may be translated to
an age difference of the order of 10 Gyr, meaning that bars may be, at least in some cases, long
standing structures. Moreover, our results show that, on average, evolved bars are
longer than young bars. This seems to indicate that, during its evolution,
a bar grows longer by capturing stars from the disk, in agreement with
recent numerical and analytical results. Although the statistical significance of these results is low,
and further studies are needed to confirm them, we discuss the implications from our results on the
possibility of bars being a recurrent phenomenon. We also present isophotal contours for all our images
as well as radial profiles of relevant photometric and geometric parameters.
\end{abstract}


\section{Introduction}

The formation and evolution of galaxies remain as a major problem in the 21st century astrophysical
sciences. Considering the complexity of this theme it could not be different, even facing the
tremendous progress done so far. It involves intricate
dynamical processes between several different components of the universe: dark matter, gas and stars.
And the proper understanding of the behavior of these components invokes more physical processes,
not all of them quite well understood.

The years that marked the demise
of the 20th century were also a period that brought a change of perspective in the way this problem
is tackled. Manifest in previous thoughts is the idea that the observed properties of galaxies were mostly
set up during violent and fast processes in the early stages of galaxy formation. This is a common
characteristic in both the monolithic \citep{egg62} and hierarchical \citep{sea78} scenarios of galaxy
formation. Referring to these seminal articles, one sees that amongst their main differences is the
time scale for the collapse of the protogalactic clouds and their physical status during the collapse.
Modern work, however, based both on observational and theoretical studies, forces the inclusion
of slow processes, that take place during a Hubble time of galactic evolution, so as to properly understand
the observed properties of galaxies. The bottom line is that, even being slow, but because they happen
so extensively in time, secular evolutionary processes are as important as galaxy formation processes to
determine what these stellar systems are, as well as how they reached their present physical
configuration.

This subject has been recently and beautifully reviewed by \citet{kor04}. In their thorough analysis
of many previous results, based upon both observational and theoretical studies, it is concluded
that one major consequence of secular evolution is the building of (pseudo) bulges \citep[see also][]{ath05}.
These are central galactic structures that produce light in excess of that from the exponential disk
but are {\em not} classical bulges (i.e., they are not formed through collapse or mergers). In fact,
pseudobulges keep a record on their disk origin via non-axisymmetric forces induced
by internal components, mainly bars, but also by oval disks and triaxial halos, and are generally flat
and disklike (whereas classical bulges are supposed to be spheroidal or weakly triaxial). As pointed out
by \citet{ath05} these different types of bulges can coexist in the same galaxy. The influence of bars
in galactic evolution, as \citet{kor04} show, has many other facets, including the building of a reservoir to
fuel the active nuclei of galaxies.

On the other hand, theories of bar formation and evolution are still in full development as the problem
posed by asking how bars form and evolve has led us to many interesting difficulties
\citep[see, e.g.,][and references therein; see also \citealt{gad03b}]{gad03a}.
Recently, \citet{ath02a} and \citet{ath02b,ath03} present promising models in the direction
of a more definite answer. These models also introduce a change of paradigm by revealing that
dark matter halos, when realistically treated, may lead to bar strengthening rather than inhibit
this instability, a belief that was spread by earlier experiments with rigid halos. Another key point
introduced by these models refers to the aging of bars. They show that bars should become longer
as they evolve. Observational tests to these models however are still lacking \citep[but see][]{lau04}.

In a companion paper \citep[hereafter Paper I]{gad04a} we introduce a new method that enables us to
discriminate between recently formed and evolved bars. Since bars are the main driver of internal secular
evolution in disk galaxies and the slow building of their central components, the next logical step is
evidently to explore links between the ages of bars and the global physical properties of galaxies.
In this paper, we proceed along this course
with the help of multiband optical and near infrared CCD images of a sample
of 19 galaxies, of which 7 have bar age estimates. In the next section we describe the sample and the
acquisition and treatment of the images. Section 3 is devoted to explaining how we have obtained
the relevant physical parameters for the discussion we address further on, that is, how we estimated
bar lengths and, from the color information in the multiband images,
global color gradients in our sample galaxies and the color of their bars.
In \S\,4 we explore these data and finally, in \S\,5, we
address a general discussion and the implications from our results on the current knowledge
concerning bar formation and secular evolution. The data presented in this paper will be used
in a future paper where we will apply a multi-component image decomposition code to explore
possible relationships between the structural properties of bulges, disks and bars.
We used a value for the Hubble constant of H$_0=70$ Km s$^{-1}$ Mpc$^{-1}$.

\section{Data Acquisition}

This section describes the main properties of our sample galaxies as well as details in the acquisition
and treatment of the optical and near infrared CCD images. For the sake of clarity this is done in
separate subsections.

\subsection{The Sample}

The more relevant properties of the 19 galaxies of our sample are displayed in Table 1. All of them
were observed in B, V, R and I, except NGC 4593, for that we could not acquire B images. For 6 of them
we also obtained images in the near infrared Ks broad band. One may easily verify that this sample
has general characteristics similar to that in Paper I; indeed both samples have 7 galaxies in common,
and for those we have also bar age estimates. We have not
looked for a complete, unbiased sample, but instead we chose interesting galaxies that could shed some
light on the subjects we are discussing.
While all galaxies are local, bright and close to face--on, they span
a range in morphologies. According to the RC3 \citep{dev91}, there is 1 SA (unbarred galaxy),
5 SAB (weakly barred) and 13 SB (strongly barred). Eight galaxies have morphological types S0 or S0/a,
4 are Sa or Sab, and the remaining 7 go as late--type as Sbc. Table 1 also shows that in our sample
there is a significant fraction of galaxies that have nuclei with non--stellar activity and/or an
identified companion that may be gravitationally interacting.

This variety might be helpful in trying to evaluate
clues related, for instance, to the prominence of the bulge, the bar length and the gravitational
perturbation of a companion. The presence of galaxies with active nuclei will also be relevant to help in
understanding the role played by bars in the fueling of this phenomenon.

The choice for bright and face--on galaxies assures us more reliable estimates for the structural parameters
of these galaxies, since it means higher signal--to--noise ratio and spatial resolution, and since
in more edge--on systems a proper description of, e.g., the bar may be unattainable. Face--on systems
also ease the interpretation of colors and color gradients as the effects of dust are minimized.

\subsection{Acquisition and Treatment of the Optical Images}

The optical imaging was done in the B, V, R and I broad bands in February and May, 1999, and in
February and March, 2000, in a total of 9 observing nights. It was performed at the Kuiper 1.55 m
telescope operated by the University of Arizona Steward Observatory on Mount Bigelow. We used
its Cassegrain f/13.5 focus and the back illuminated CCD no. 24, with a gain of 3.5 e$^-$
ADU$^{-1}$ and a readout noise of 8.7 e$^-$. This CCD has 2048 $\times$ 2048 21 $\mu m$ pixels
but we used it binned in 2 per 2 pixels. Thus, the plate scale of these observations is 0.29'' per
pixel, and their field of view is roughly about 5' on a side. The filters transmission curves
are similar to the Johnson--Morgan system for B and V, and to the Cousins system for
R and I \citep[see, e.g.,][]{fuk95,kit98}. Table 2 shows a summary night log with relevant information.

To ease cosmic ray removal we have taken
several exposures for each galaxy. These were 5 in B and V, and 3 in R and I with an integration
time of 300 seconds each. Standard stars from \citet{lan83} were used for photometric calibration.
This was done in the standard manner within the {\sc iraf}\footnote{{\sc iraf} is distributed by the National
Optical Astronomy Observatories, which are operated by the Association of Universities for Research in
Astronomy, Inc., under cooperative agreement with the National Science Foundation.} environment.
Table 2 also shows a summary of these calibrations.
To correct extinction caused by the atmosphere, appropriate coefficients were determined for each band.
Note that the photometric errors are larger in the I band, which is expected as this band covers sky emission
lines and its images had to be corrected from fringing effects.

Pre-processing of the images for analysis was also performed with {\sc iraf}.
Bias images were subtracted from all science images and the latter were normalized with dome flatfields
in the usual manner. Fringes in the I band images were removed with the {\sc mkfringecor} task.
Finally, bad pixels were also eliminated but the detectors used were cosmetically clean.

To eliminate cosmic rays and raise the final S/N in the science images we used the {\sc imcombine} task to
combine by the median multiple exposures of each galaxy in each band. Small shifts between the exposures
were calculated with the {\sc imcentroid} task and 3 to 5 reference stars.

To remove the contribution from the sky we first edited the combined
images, removing the galaxy and foreground stars. This was done using the task {\sc imedit}
with a circular aperture and replacing the selected pixels by background pixels with a Gaussian
noise distribution added. After that, we
determined the mean sky background and its standard deviation. Then,
we removed all pixels whose values were discrepant from
the mean background by more than 3 times the
standard deviation. A sky model was obtained by fitting with {\sc imsurfit} a linear surface to the
image, and this model was subtracted from the combined image. We finally removed
objects such as foreground stars and bright H{\sc ii} regions from the sky subtracted image to get a final
science image for each galaxy in B, V, R and I. It should be noted, however, that for 4 galaxies
(NGC 3227, 4151, 4303, and NGC 4579) our images may not contain areas in which there is only a background
contribution. This means that the sky contribution may have been overestimated. Hence, the faintest parts
of their optical surface brightness profiles may be subject to this source of error.

\subsection{Acquisition and Treatment of the Near Infrared Images}

Images in the near infrared Ks band were acquired for 6 galaxies in our sample with the 2.29 m Bok telescope
operated by the University of Arizona Steward Observatory on Kitt Peak. Observations occurred in 5 nights
during May and June 1999. The focal ratio in the Cassegrain mode was f/45. We used a Rockwell infrared
detector with $256^2$ pixels, with a readout noise of 30 e$^-$ and gain of 15 e$^-$ ADU$^{-1}$.
The pixel size on sky was 0.59 arcsec and the field of view nearly 2.5 arcmin. The filter's transmission
curve is approximately flat from 2.0 $\mu$m to 2.3 $\mu$m with sharp cutoffs. Table 3 shows relevant
data on the near infrared observations and photometric calibrations.

Our observing strategy in the Ks images was typically to obtain two exposures of 30 s in the galaxy,
followed by two sky exposures of 30 s in regions relatively free of objects a few arcmin away from the galaxy,
that serve for sky subtraction. Such a cycle is repeated several times in a square dither pattern with shifts
of several arcsec. These procedures avoid space and time sky changes and the effects of bad pixels;
also, as the field of view is restricted, it allows us to cover a larger
area of the galaxy than with just one pointing. In those galaxies with an angular apparent diameter larger than
our images the loss is restricted to the disk outskirts and does not compromise our analysis. In particular,
the bar is always entirely visible.

For the photometric calibration we observed standard stars from \citet{eli82} and \citet{per98}, and
determined an atmospheric extinction correction coefficient. As this calibration in the near infrared
is prone to more errors than in the optical case we observed many standard stars and most of them at
several zenithal distances. Table 3 displays the photometric errors calculated in the
nights they were observed.

In the treatment of the near infrared images we used the {\sc gemini} {\sc iraf} package. Flatfield images
were obtained through the {\sc qflat} task from all sky images related to a determined galaxy. The {\sc qsky}
task was used to estimate the sky contribution from each cycle. These corrections were performed by
{\sc qreduce}. Finally, the {\sc imcoadd} task combines by the median all corrected images of a galaxy
calculating the necessary shifts due to the dither pattern, and also corrects bad pixels.

\section{Obtaining Relevant Physical Parameters}

\subsection{Bar Lengths}

Using our images and the {\sc ellipse} task in {\sc iraf} we built for each galaxy in each band
radial profiles of the elliptically averaged surface brightness, as well as geometric parameters of
the isophotes, namely, position angle, ellipticity and the b4 Fourier coefficient. The results are shown
in Fig. 1, that also displays our images and isophotal contours. During the ellipse fitting the center was held
fixed. To find the galaxy center we first run {\sc ellipse} with the center free and chose the center
as the average of the values chosen by the task at a radius around 10 to 15 pixels. The radius for each
fit grows logarithmically at a 10\% rate.

The optical surface brightness profiles were corrected from the extinction
by the Galaxy following the maps of \citet{sch98} in the NASA Extragalactic Database
(hereafter NED). For the sake of precision, we applied a first order correction for the dust extinction
caused within the galaxies themselves;
we used the results from \citet{gio94} that find a relation between the galaxy inclination and the
extinction, that may be expressed in I magnitudes as

\begin{equation}
{{A}_{I}} = {1.12(\pm 0.05) \log{\rm R}_{25}}.
\end{equation}

\noindent To find the analogous equations for the other optical bands one just needs to apply the
results of \citet{elm98} that shows that the extinction in B, V and R is, respectively, 3.17, 2.38 and 1.52
times larger than in I.

We next tried to estimate the lengths of the bars, $L_B$, in the galaxies
of our sample in a most reliable fashion as possible.
This was done by calculating the average radius where the end of the bar is detected in the
V and R images through 3 different criteria: a peak in the ellipticity radial profile, the mean point of
an abrupt change in position
angle (in the cases where the position angle of the bar is different from that of the disk), and the point where
the galaxy surface brightness profile drops to join back the light distribution in the disk
-- in the bar region there is of course an excess of light if one compares it to the profile of an unbarred
galaxy or the exponential profile of the disk (see Fig. 1).
We also noticed that near the end of the bar, but somewhat closer to the center, there is a peak in the
b4 Fourier component, which is expected due to the non-axisymmetric nature of the bar. We used only the V
and R images in this analysis aiming to obtain a compromise between S/N and a better representation
of the majority of the stellar population in the galaxies, but we did verify that no discrepancies would arise
if other band images were used instead.

Since most of our galaxies are close to face--on, projection effects
on the bar length estimates are negligible. However, five barred galaxies in our sample may be somewhat inclined systems
as their log R$_{25}$ parameter is larger than 0.1 (and assuming that their outermost isophotes are intrinsically
circular -- see Table 1). Their projected (observed and measured)
bar lengths thus may be, in principle, underestimated if compared to the real lengths.
Fortunately, however, in each of these galaxies the position angle of the bar is very similar to that of the line
of nodes (the largest misalignment is $\sim20^\circ$), meaning that the bar length would
not change significantly after deprojecting the galaxy image. We made this analysis using our data
as well as that present in LEDA and in NED and found an excellent agreement.
Obtaining deprojected measurements for the bar lengths could also introduce
errors from the uncertainties in the inclination angle and the position angle of the line of nodes.
To consider that the outermost isophotes of disk galaxies are intrinsically circular may not be a safe
assumption. In addition, deprojection techniques assume that the whole galaxy is flat, which is not
the case of evolved, vertically thickened bars.
We believe that taking projection effects into account would only prove useful (rather than harmful)
if most of the galaxies in our sample were far from face--on and if our sample was sufficiently larger to significantly
put down statistical random errors in the measurements of $L_B$ itself.
A discussion on the advantages and disadvantages of deprojecting galaxy images can also be found in
\citet{jun97} and in \citet{erw03}.
Furthermore, since none of these more inclined galaxies in our sample were studied in Paper I, and then
had the age of their bars evaluated, the main results of this paper would remain unchanged even after
considering projection effects.

We looked for a correlation between $L_B$ and D$_{25}$ from our B images (and from the RC3
when necessary, i.e., for NGC 4303, NGC 4593 and NGC 5850 -- in the remaining cases our values agree well
with those from RC3) and a trend was found but with a large spread
(the linear correlation coefficient is 0.5, see Fig. 2). This means that it is not necessarily true that
bigger galaxies have bigger bars, anticipating that $L_B$ might also be ruled by another physical process(es).
Table 4 shows the results. It is interesting to note, however, that \citet{lai02} found a somewhat better
correlation (their linear correlation coefficient is 0.66). Recently, \citet{erw05} found also a better
correlation. Nonetheless, his results show that the correlation is stronger for S0 galaxies and weakens
significantly towards later morphological types. In fact, he found that bars in S0--Sab galaxies extend to
$\sim 0.2-0.8 R_{25}$ whereas late--type bars extend to only $\sim 0.05-0.35 R_{25}$. These figures also
reflect the significant fluctuations in the correlation.

Measuring bar lengths has ever proved a task harder than it seems. Thus a comparison of our results to those
already published might be useful. For 8 galaxies of our sample a method was already applied to estimate the
bar length, and in some cases more than one method. \citet{mar95a} made visual measurements of $L_B$ on
photographic plates of good linear scale for 3 of these galaxies: NGC 4303, where his estimates point to a
value of 2.2 Kpc whereas ours give 3.2 Kpc; NGC 4394, again with estimates shorter than ours, 2.5 Kpc against
3.7 Kpc; and NGC 5850 where, remarkably, our estimates are identical, 10.6 Kpc. In the next studies we will
discuss, bar lengths were also measured through the analysis of geometrical properties of the isophotes but
the choice for radius is different. \citet{woz95} chose the bar radius as the one where the ellipticity drops
to a local minimum after the maximum reached in the bar. They measured $L_B$ for NGC 4593 as 10.4 Kpc, whereas
we found 10.8 Kpc. They also give $L_B$ for NGC 5850: 14.8 Kpc, considerably in excess of the values above.
The bar in the latter has also been measured by \citet{erw04} who arrived to the intermediate value of 13.7 Kpc.
His choice for the bar radius is the minimum between the radius of the local minimum in ellipticity (as in
\citet{woz95}) and the radius, after the ellipticity peak in the bar, where the position angle changes by at
least 10 degrees. He also presents estimates for $L_B$ in NGC 4303 (3.8 Kpc whereas ours is 3.2 Kpc) and NGC
4314 (6.3 Kpc while ours is 7.4 Kpc). Using similar criteria, \citet{erw02,erw03} give $L_B$ for NGC 4665 as
2.7 Kpc, while we found 3.2 Kpc. Finally, \citet{jun97} discuss bar lengths as the point where the ellipticity
peaks, finding for NGC 4267 an $L_B$ of 1.5 Kpc (we found 1.7 Kpc) and for NGC 5701 an $L_B$ of 4.2 Kpc (while
our result is 4.5 Kpc). The mean difference between our estimates and those in the literature is about 19\%.
Altogether, one sees general agreement, but the discrepancies mean that the task of finding a better
definition for the bar length is certainly one worthy to pursue.

\subsection{Bar Colors and Galaxy Global Color Gradients}

In Fig. 3 we display optical color profiles for the galaxies in our sample, whereas Fig. 4 presents
B-R color maps (except for NGC 4593 where the color map refers to V-R).
The color profiles were built from the ellipse fits. Although the set of ellipses used are not exactly
the same in each band, the relevant geometrical properties of the ellipses are generally identical in 
different bands, and thus light from different parts of the galaxy in different bands is not mixed. 
From the color profiles we have estimated B-V, B-R and B-I color gradients for all galaxies, except for
NGC 4593, whose color gradients refer to V-R and V-I. The gradients were calculated through linear
regression and the least squares method, excluding the nuclear region affected by seeing. Moreover,
we have also eliminated outlier points by visual inspection to avoid spurious results, but this scarcely happened.
Thus, these color gradients represent a global galaxy tendency. The gradients are expressed as in
\citet{gad01}, i.e., $G(X-Y) \equiv \Delta (X-Y)/\Delta \log A$, where $X-Y$ is the color index and
$A$ is the isophote diameter in units of 0.1 arcmin. Using these profiles we also estimated
the B-I color of the bars, $(B-I)_B$.
These were obtained at the point that marks the end of the bar, i.e., $L_B$, but
we have also avoided star forming sites that could lead to misleading results.
The choice for the B-I index is discussed below. All these parameters
are shown in Table 4. These color values are all corrected for dust reddening in the Galaxy, and inclination
corrections were also applied (again, for the sake of precision).
We used the same correction as for the surface brightness, except,
of course, that we now used the color excess $E(X-Y)=A_X-A_Y$.

The correlations between the different color gradients are shown in Fig. 5. As was already discussed
in other studies \citep[e.g.,][]{dej96b,gad01} these correlations show that the same physical process
is responsible for creating these gradients. These color gradients could be a result of either metallicity, age
or dust effects, but \citet{gad01} argue that, at least for late--type spirals (around
type Sbc), it is the age of the stellar population that is mostly responsible for the global color
gradients. From the plots one may linearly fit

\begin{equation}
G(B-R) = -0.05(\pm0.02)+1.00(\pm0.12)\times G(B-V)
\end{equation}

\noindent and

\begin{equation}
G(B-I) = -0.01(\pm0.05)+1.80(\pm0.27)\times G(B-V),
\end{equation}

\noindent which means that $G(B-V)$ and $G(B-R)$ are almost equivalent, and suggests that $G(B-I)$
may be more sensitive to variations in the stellar population ages, since the slope in Eq. (3) is steeper.
Note, however, that the scatter in the $G(B-I)$/$G(B-V)$ relation is also larger; but since the overall
trend is similar in both plots of Fig. 5 the results concerning bar colors should in principle not
depend significantly on the choice of color index.

\section{Analysis}

\subsection{Does Bar Color Give Bar Age?} 

From Table 4 one can see that the mean value for the bar color index
$(B-I)_B$ is 1.79$\pm$0.11.\footnote{Errors quoted are the standard error on the mean, corresponding to
1-$\sigma$ confidence level.}
In Paper I we show that judging from the kinematical analysis NGC 4314, 4608, 5701 and NGC 5850 have evolved bars,
whereas those in NGC 4394, 4579 and NGC 5383 are recently formed. The average $(B-I)_B$ of the former
is 2.17$\pm$0.12 while that of the latter is 1.49$\pm$0.20. Old bars are on average
0.68 mag redder than young ones, and this difference is significant at the
3-$\sigma$ confidence level.\footnote{Even so, this result should be carefully considered since with the
small number of galaxies in this analysis the mean and the errors are poorly determined.} Obviously, the color
of the bars reflect not their ages but the ages of their stars. However, barred galaxies follow
two different patterns regarding star formation \citep{phi93,phi96,mar97}. Later--type spirals
show star forming sites {\em along} their bars, that are shorter and have an exponential luminosity
profile \citep{elm85}. Earlier--type spirals, with longer bars that have flat luminosity profiles, lack any star
formation along their bars, but present star forming sites in their nuclear and inner rings.
\citet{mar95b} and \citet{fri95} argue that star formation along the bars could indicate that these
structures are young ($\lesssim 1$ Gyr). Thus, the age of the stellar population within the bar may
be an indication of the age of the bar itself. This is also suggested by our results that the evolved bars
from Paper I indeed have redder colors than the recently formed bars. Nevertheless, we are neglecting
effects from differential dust extinction along the bar and the age--metallicity degeneracy, but this
seems to be fairly justified since $(B-I)_B$ is measured very far from the centers of the galaxies
(where dust extinction is more pronounced), and considering the results from \citet[see also
\citealt{dej96b}]{gad01} that global color gradients are not affected by dust and are mainly
sensible to the age, rather than to the metallicity, of the stellar population. In addition, since
the galaxies in our sample are in general close to face--on dust extinction is further minimized.

The average B-V color of the bars that the results from Paper I indicate as evolved bars is 1.1,
whereas that of the young bars is 0.7. Theoretical studies of the evolution of the stellar population
in galaxies \citep[e.g.,][]{tin76,mar98} indicate that the mean age of the stars in the
latter case is about 1 to 2 Gyr, while
the mean age of the stellar population in evolved bars would be at the order of 15 to 20 Gyr. Although
these results are model dependent a reasonable difference in age from young and evolved bars seems to be
of the order of 10 Gyr. This result reinforces those by \citet{she04} that indicate that bars
are a robust structure \citep[see also][]{elm04,jog04b}, and also our results from Paper I in which we
suggest a slow mechanism for the
vertical evolution of bars, namely the Spitzer--Schwarzchild mechanism \citep{spi51,spi53}.

\subsection{Do Bars Grow Longer While Aging?}

The recent theoretical studies by \citet{ath02a} and \citet{ath02b,ath03} suggest that during
its evolution bars grow longer by capturing stars from the disk and redistributing angular momentum
along the disk and halo.
In Paper I we show that while NGC 4314, 4608, 5701 and NGC 5850 have evolved bars the opposite
is true for NGC 4394, NGC 4579 and NGC 5383, with recently formed bars. In agreement
with the picture suggested by these theoretical models,
the evolved bars are on average 2.1 Kpc longer: The average $L_B$ for evolved bars is 7.5$\pm$1.2 Kpc
and for young bars is 5.4$\pm$1.6 Kpc. The average value of $L_B$ for the whole sample is 5.0$\pm$0.7 Kpc.
Similar results are obtained if one uses the length of the bar normalized
by $D_{25}$.

It should be stressed that our sample is statistically small. The difference between $L_B$ for young and old
bars is reliable only at the 1-$\sigma$ confidence level, and thus it is necessary to obtain bar age
estimates for a larger galaxy sample. On the other hand, the length of the
bar does not depend only on its age and the disk size.
\citet{ath03} shows that bar strength depends on a series of factors, notably the ability of the
disk and halo to exchange angular momentum via orbital resonances. This may partially explain the spread
in our age $\times L_B$ relation.

\subsection{Bar Profiles}

It is interesting to ask whether young and old bars have also different luminosity profiles
along their major and minor axes. In Fig. 6 we plot these profiles separately for young and evolved
bars and see no difference. All galaxies show luminosity profiles along their bar major axis
with conspicuous plateaus, while their bar minor axis profiles are closer to exponential.
As shown by \citet{elm85} this is characteristic of bars in galaxies with morphological
types earlier than Sbc.

\section{Discussion}

The relation between bar color and age must be treated with care and at least two caveats kept
in mind. First, we showed that young bars look bluer than old ones, but since young bars seem to be
found mostly in late--type galaxies it must be checked whether this difference in bar color
is not a result of the difference in the color of the galaxies as a whole, as late--type galaxies
have generally integrated colors bluer than early--type galaxies \citep{rob94}, nevertheless
with a large spread. Second, it is not clear how the color of a bar would change when it captures
stars from the disk as it evolves. In principle, one may argue that the capture of stars from the disk
would tend to avoid it getting redder, and thus dilute the relation between bar age and color.

Table 5 shows a summary of the average properties of the bars in our sample concerning the
galaxies' morphological types and the prominence of the bar according to the RC3. As expected,
weak bars (in SAB galaxies) are shorter than the strong ones (SB). They are also bluer on average, as
are bars in later--type galaxies. It appears, however, that it could be not the case only that bluer
bars reside in bluer galaxies and vice--versa. That would represent a bias towards pointing to
younger bars in later--type galaxies. To assess this issue we can take from LEDA integrated B-V colors (corrected
to face--on) of those galaxies for which we have bar age estimates and compare them. The mean global B-V
color of galaxies with old bars is 0.81$\pm$0.05 whereas that of galaxies with young bars is 0.72$\pm$0.06.
Although there seem to be a difference (0.09 mag) it appears too small to explain the difference we
found in the B-V color of young and old bars, which is of 0.40 magnitudes. Hence, the result on the bluer
bars of SAB galaxies may be genuine and point to a very interesting conclusion. SAB bars may be the ones
which are passing through processes of formation, or, alternatively, dissolution. However, their bluer
colors suggest they are forming, and thus young, bars. Furthermore, either the processes of bar dissolution
are significantly faster than those of bar formation, or bar dissolution is much less frequent than bar formation,
since, otherwise, those dissolving bars, with their old and red stellar population, would dilute this difference
in color.

One may also argue that the capture of disk stars by the bar does not turn it
bluer. Firstly because the bar originates from the same stellar population that makes up the
original disk, and, secondly, because after the star forming bursts along the recently formed bar fade
new stars will mostly appear within the disk only outside of the bar region. Gas in the disk is collected
by the bar and funnelled away from the bar region. It is evident that further
theoretical and observational studies on the evolution of bars and its effects on the
overall star formation
in galaxies are necessary to clarify these issues, as is a large sample of galaxies with
good estimates for bar ages, colors and lengths.

In Paper I \citep[see also][]{gad04b} we find that galaxies with AGN tend to have young bars while
galaxies with old bars do not generally have AGN. This leads us to conclude that the fueling
of AGN by bars occurs in short time scales, i.e., of the order of 1 Gyr or less. Here we also find
results that corroborate this conclusion if we assume the B-I color of the bars as a good
index of the bars' ages. In fact, the B-I color of the bars in galaxies with AGN in our sample is 1.63$\pm$0.11,
whereas this goes to 2.03$\pm$0.19 in galaxies without AGN. Hence, it appears that the building of a gas reservoir
for AGN by a bar occurs in the bar first evolutionary stages. As galaxies with AGN are generally early--type
it is not likely that this difference in bar color is a result from the difference in color of the galaxies
as a whole, as the trend would thus be opposite.

\citet[and references therein]{lau04} show that the bars in galaxies with AGN while being massive may
have a weaker impact on the overall galaxy evolution when compared to galaxies without AGN. Considering that the amount
of fuel necessary to ignite an AGN episode is {\em very} small \citep[see, e.g.,][]{jog04a} this
result is not as surprising as it may appear.
In this context it is worth stressing here that when comparing different bars the fact that one is
longer does not necessarily means that it produces greater changes
in the galaxy mass distribution. As shown by \citet{lau04} if there is a
prominent bulge its effects on the force distribution along the galaxy will dilute those effects
caused by the bar. However, when one says that a bar grows longer, as in Athanassoula's models,
it is obvious that it also grows stronger, as it gets more massive, and possibly more eccentric,
unless the bulge also grows more important accordingly.

Another relevant question concerns the possibility of recurrent bars. In this regard, it is interesting
that from the spectroscopy results in Paper I we find amongst evolved bars 7/8 galaxies with
morphological types S0 to Sa, and only 1/8 Sb galaxy. On the other hand, from the 5 galaxies with young bars
2 are S0 to Sa, and 3 are Sb galaxies. This is interesting considering the suggestion by \citet{fri95}
that bars are young in late--type galaxies and more evolved in early--type galaxies. This is also consistent
with our conclusion that SAB galaxies have young bars, as we find in \citet{gad04b} that SAB galaxies
occur preferentially in the Sc bin of the morphological classification scheme. Unless bars are recurrent
and contribute significantly
to bulge building only in late--type galaxies these findings challenge bar recurrent scenarios such as the
one by \citet{bou02}. If bars are recurrent and
contribute to the formation of bulges along the Hubble sequence we should
not observe a higher fraction of young bars in one narrow range of morphological types.
In the \citet{bou02} scenario bars die and reborn with ever shorter lengths and faster pattern
rotation velocities, $\Omega_B$. If bars are not recurrent, however, and grow longer while aging,
then we should expect $\Omega_B$ to fall in early--type galaxies. The results concerning this physical
parameter are still very rare (especially in late--type galaxies), but initially point to lower values
for $\Omega_B$ in early--type galaxies \citep{ger03}.
Thus it seems that the models of \citet{bou02} are not consistent
with the observations. In the recurrent models of \citet{ber04} bars reappear with a lower rotation
velocity but these results only apply for galaxies that lack gas and when bars are a result from
tidal forces provoked by a close companion. Another argument against the \citet{bou02} recurrent bar
scenario, based directly on bar sizes, has recently been made by \citet{erw05}.

The correlation found by \citet{ath80} in that more prominent bulges appear in galaxies with
longer bars is also an argument that favors a scenario in which as bars age and grow longer
they contribute secularly to bulge building. Furthermore, if this picture is correct, then we
should not expect that bars are recurrent, since in this case we should observe young, short
bars in galaxies with big bulges, which is not the case. Hence, if bar formation processes are similar
in early-- and late--type galaxies, it seems that either bars are recurrent
only in galaxies with morphological types around Sc, or are generally not recurrent at all.

\acknowledgments
We thank Rob Kennicutt for his kind hospitality and for giving us access to the Steward
Observatory telescopes. It is a pleasure
to thank Lia Athanassoula for her comments on a first draft of this article. We are thankful to the
anonymous referee for a careful reading of the manuscript and for many useful remarks.
This work was financially supported by FAPESP grants 99/07492-7 and 00/06695-0.
This research has made use of the
NASA/IPAC Extragalactic Database (NED), which is operated by the Jet Propulsion
Laboratory, California Institute of Technology, under contract with the National Aeronautics and Space
Administration. This research has also made use of NASA's Astrophysics Data System and of the HyperLeda
database (\url{http://leda.univ-lyon1.fr/}).

\newpage

\figcaption[f1*.*ps]{Results of the ellipse fitting to the optical and near infrared images of all galaxies
in our sample. Each CCD image is median-filtered and has isophotal contours overlaid, spaced by 0.5
mag arcsec$^{-2}$. In the optical bands the images refer to B, V, R and I broadbands from left to right and
from top to bottom. The radial profiles are of surface brightness (top left), position angle (bottom left),
ellipticity (top right) and the b4 Fourier coefficient (bottom right). Position angles grow from North at
0$^\circ$ through East at 90$^\circ$. North is up and East to the left in all optical images. In the near
infrared images North is up and East to right. The remaining figures are available in the electronic edition of the Journal.
The printed edition contains only a sample.}

\figcaption[f2.eps]{The radius of the 25 B mag arcsec$^{-2}$ isophote plotted against the length of the
bar, i.e., its semi-major axis.}

\figcaption[f3*.eps]{Color profiles for all galaxies in our sample. The solid line refers to B-V, the dotted
one to B-R and the dashed line to B-I. The vertical solid line indicates the typical seeing while
the dotted one is our estimate for $L_B$, from where the color of the bar is derived. Color units
are magnitudes and the ordinate is the galactocentric radius expressed in units of arcsec in a logarithmic
scale.}

\figcaption[f4*.eps]{B-R color maps for the galaxies in our sample, except for NGC 4593 where it is V-R.
Darker areas indicate bluer colors.}

\figcaption[f5.eps]{$G(B-R)$ and $G(B-I)$ plotted against $G(B-V)$. The solid lines are linear fits to the
data in each panel.}

\figcaption[f6.eps]{R-band profiles of the bar major and minor axes for recently formed and evolved bars.}

\newpage
\begin{deluxetable}{llccccccc}
\tablecaption{Basic data for all galaxies in our sample.}
\tablewidth{0pt}
\tablehead{Name & Type & D$_{25}$ & log R$_{25}$ & m$_{\rm B}$ & cz & d & AGN & Companion \\
(1) & (2) & (3) & (4) & (5) & (6) & (7) & (8) & (9)}
\startdata
IC 0486     & SBa              & 0.93         & 0.11         & 14.60          &     7792 & 111.3    & Sey1            & M     \\
NGC 2110    & SAB0           & 1.70          & 0.13         & \dots           &    2064 & 29.5    & Sey2            & N     \\
NGC 2493    & SB0             & 1.95          & 0.00         & 12.91          &    4060 & 58.0     & \dots            & N     \\
NGC 2911    & SA0(s)         & 4.07          & 0.11         & 12.21          &    3195 & 45.6     & Sey/LINER     & Y     \\
NGC 3227    & SABa(s)       & 5.37          & 0.17         & 11.59          &    1235 & 17.6     & Sey1.5         & Y     \\
NGC 4151    & SABab(rs)    & 6.31          & 0.15         & 10.90          &     1190 & 17.0     & Sey1.5          & M     \\
NGC 4267    & SB0(s)         & 3.24          & 0.03         & 11.73          &    1123 & 16.0     & \dots            & N     \\
NGC 4303    & SABbc(rs)    & 6.46          & 0.05         & 10.21          &    1620 & 23.1     & Sey2            & Y     \\
NGC 4314$^b$    & SBa(rs)        & 4.17          & 0.05         & 11.22          &     1146 & 16.4      & LINER            & N     \\
NGC 4394$^{a,b}$    & SBb(r)          & 3.63          & 0.05         & 11.53          &     1036 & 14.8      & LINER            & Y     \\
NGC 4477    & SB0(s)         & 3.80           & 0.04         & 11.27          &    1441 & 20.6     & Sey2            & Y     \\
NGC 4579$^b$    & SABb(rs)      & 5.89          & 0.10         & 10.68          &    1607 & 23.0     & LINER/Sey1.9 & N     \\
NGC 4593    & SBb(rs)        & 3.89           & 0.13         & 11.67          &    2498 & 35.7     & Sey1            & M     \\
NGC 4608$^{a,b}$    & SB0(r)          & 3.24          & 0.08         & 11.96          &    1893 & 27.0     & \dots            & N     \\
NGC 4665    & SB0/a(s)      & 3.80          & 0.08         & 11.50          &     872 & 12.5      & \dots            & N      \\
NGC 5383$^{a,b}$  & SBb(rs)        & 3.16          & 0.07         & 12.18          &    2472 & 35.3     & \dots            & Y      \\
NGC 5701$^{a,b}$    & SB0/a(rs)     & 4.26           & 0.02         & 11.82          &    1601 & 22.9     & LINER            & N      \\
NGC 5850$^{a,b}$    & SBb(r)          & 4.26           & 0.06         & 12.04          &    2637 & 37.7     & \dots            & N     \\
NGC 5936$^a$    & SBb(rs)        & 1.44           & 0.05         & 13.01          &    4147 & 59.2     & \dots            & N      \\
\enddata
\tablecomments{Columns (1) and (2) show, respectively, the name and the morphological type of the
galaxy, while column (3) shows its diameter in arcminutes at the 25 B magnitude isophotal level, and
column (4) shows the decimal logarithm of its major to minor axes ratio at the same level. Columns (5)
and (6) show, respectively, the apparent B magnitude and the radial velocity in Km/s. All these data were
taken from \citet[hereafter RC3]{dev91}, except the radial velocity, taken from the Lyon Extragalactic Data
Archive (herefater LEDA), corrected for infall of the Local Group towards Virgo. Column (7) gives the distance
to the galaxy in Mpc, using the radial velocity in column (6) and H$_0=70$ Km s$^{-1}$ Mpc$^{-1}$.
Column (8) presents an Active Galactic Nuclei classification according
to the NASA Extragalactic Database (hereafter NED). In column (9), ``Y'' means that there is a companion
galaxy similar in size physically interacting within 10 arcminutes, while ``N'' means that there are
no companion galaxies; ``M'' indicates that there are companion galaxies that lack redshift data and
thus may be only projection effects. To make this analysis we used the RC3 and LEDA. $^a$ indicates those
galaxies with images also in the near infrared, and
$^b$ those in common with Paper I that have also bar age estimates.}
\end{deluxetable}

\newpage

\begin{deluxetable}{llcccccc}
\rotate
\tablecaption{Summary of the optical observations and calibrations.}
\tablewidth{0pt}
\tablehead{Night & Galaxy & Seeing (arcsec) & Photometric? & Error (B) & Error (V) & Error (R) & Error (I)}
\startdata
17/Feb/99 & N2493 & 1.4 & no & \omit & \omit & \omit & \omit \\
09/May/99 & N4680;N5701;N5936BVR & 1.4 & yes & 0.01 & 0.01 & 0.01 & 0.01 \\
10/May/99 & N4267;N4665;N5850 & 1.3 & no & \omit & \omit & \omit & \omit \\
11/May/99 & N4394;N4477;N5936I & 1.2 & yes & 0.04 & 0.02 & 0.01 & 0.35 \\
12/May/99 & N4314;N5383 & 1.2 & yes & 0.05 & 0.01 & 0.02 & 0.38 \\
01/Feb/00 & I0486;N3227;N4593VRI & 1.4 & yes & \omit & \omit & \omit & \omit \\
02/Feb/00 & N2110 & 1.5 & yes & 0.01 & 0.02 & 0.01 & 0.02 \\
03/Feb/00 & N2911;N4151;N4579 & 1.4 & no & \omit & \omit & \omit & \omit \\
03/Mar/00 & N4303 & 1.3 & yes & 0.01 & 0.01 & 0.01 & 0.01 \\
\enddata
\tablecomments{Galaxies observed in each usable night of the observing runs. When no
band is specified the observations were done in B, V, R and I. The average seeing in the R band
in arcseconds and the photometric conditions are also displayed, along with the photometric
zero point error from standard calibration in each band. In the nights when no
standard stars were observed the zero point of the closest photometric night was applied.}
\end{deluxetable}

\newpage

\begin{deluxetable}{llcccc}
\tablecaption{Summary of the near infrared observations and calibrations.}
\tablewidth{0pt}
\tablehead{Night & Galaxy & Seeing (arcsec) & Photometric? & Error (Ks) & t (s)}
\startdata
04/May/99 & N4608;N5850       & 1.2 & yes & 0.03    & 570;900 \\
05/May/99 & N5701             & 1.3 & yes & 0.09    & 1350 \\
29/May/99 & N5936             & 1.2 & no  & \omit   & 180 \\
31/May/99 & N5383             & 1.1 & yes & 0.03    & 840 \\
01/Jun/99 & N4394             & 1.2 & yes & \omit   & 230 \\
\enddata
\tablecomments{Galaxies observed in each usable night of the observing runs.
The average seeing in arcseconds and the photometric conditions and errors after
standard calibration are also displayed.
The total exposure time for each galaxy is in seconds in the last column. In the nights when no
standard stars were observed the zero point of the closest photometric night was applied.}
\end{deluxetable}

\newpage

\begin{deluxetable}{lcccccc}
\tablecaption{Global color gradients and bar color and length.}
\tablewidth{0pt}
\tablehead{Galaxy  &       $G(B-V)$   &       $G(B-R)$  &        $G(B-I)$   &      $(B-I)_B$ & $L_B$ (Kpc) & $2L_B/D_{25}$}
\startdata
I486        &    -0.17$\pm$0.01 &  -0.17$\pm$0.01 &  -0.34$\pm$0.01       &        1.44     &  4.7    &     0.30               \\
N2110     &   -0.08$\pm$0.01  &  -0.13$\pm$0.01 &  -0.11$\pm$0.01       &       1.46      &  3.3    &     0.31              \\
N2493     &   -0.19$\pm$0.01  &  -0.18$\pm$0.01 &  -0.21$\pm$0.01       &       2.44      &  7.7    &     0.67               \\
N2911     &    -0.09$\pm$0.01 &  -0.21$\pm$0.01 &  -0.36$\pm$0.01       &        \dots   & \dots  &      \dots                  \\
N3227     &    0.11$\pm$0.01  &  -0.16$\pm$0.02 &  -0.09$\pm$0.02       &       1.71      &  1.9    &     0.16                 \\
N4151     &    0.25$\pm$0.01  &   0.31$\pm$0.02 &   0.63$\pm$0.03        &      1.62       &  4.5    &     0.67                   \\
N4267     &    0.04$\pm$0.00  &   0.08$\pm$0.00 &   0.10$\pm$0.01        &      2.48       &  1.7    &     0.26                    \\
N4303     &    -0.25$\pm$0.04 &  -0.26$\pm$0.03 &  -0.29$\pm$0.04       &       1.08      &  3.2    &     0.15                     \\
N4314     &     0.02$\pm$0.01 &  -0.08$\pm$0.02 &  -0.05$\pm$0.02       &       1.90      &  7.4    &     0.84                   \\
N4394     &    -0.09$\pm$0.01 &  -0.17$\pm$0.01 &  -0.17$\pm$0.02       &       1.30      &  3.7    &     0.56                   \\
N4477     &    -0.09$\pm$0.01 &  -0.14$\pm$0.02 &  -0.17$\pm$0.02       &       1.60      &  2.9    &     0.31                    \\
N4579     &    -0.11$\pm$0.01 &  -0.14$\pm$0.01 &  -0.13$\pm$0.01       &       1.89      &  3.9    &     0.32                    \\
N4608     &    -0.07$\pm$0.01 &  -0.02$\pm$0.01 &  -0.15$\pm$0.01       &       2.05      &  7.8    &     0.67                \\
N4665     &    -0.10$\pm$0.01 &  -0.23$\pm$0.01 &  -0.03$\pm$0.02       &       2.18      &  3.2    &     0.40                   \\
N5383     &    -0.43$\pm$0.04 &  -0.50$\pm$0.05 &  -0.57$\pm$0.07       &       1.29      &  9.7    &     0.80                    \\
N5701     &    0.02$\pm$0.00  &  -0.07$\pm$0.01 &   -0.10$\pm$0.01       &      2.32       & 4.5     &    0.47                     \\
N5850     &   -0.12$\pm$0.01  &  -0.14$\pm$0.01 &   -0.09$\pm$0.01      &       2.40      &  10.6  &     0.45                     \\
N5936     &   -0.38$\pm$0.02  &  -0.48$\pm$0.03 &   -1.14$\pm$0.04       &      1.34       & 1.6     &    0.12                  \\
\tableline
\tableline
Galaxy  &      \omit   &       $G(V-R)$  &        $G(V-I)$   &      $(V-I)_B$ & $L_B$ (Kpc)  & $2L_B/D_{25}$  \\
\tableline
N4593     &      \omit       &         -0.13$\pm$0.01  & -0.01$\pm$0.01    &       0.92     &    10.8    &     0.50                    \\
\enddata
\end{deluxetable}

\clearpage

\begin{deluxetable}{lcc}
\tablecaption{Average properties of bars. Standard error in parenthesis.}
\tablewidth{0pt}
\tablehead{Type & $L_B$ (Kpc) & $(B-I)_B$}
\startdata
0--0/a     &      4.4 (0.9)     &           2.08 (0.14) \\
a--ab      &      4.6 (1.1)    &            1.67 (0.09)   \\
b--bc      &     6.2 (1.5)    &            1.55 (0.20)      \\
\tableline
SAB   &          3.4 (0.4)   &            1.55 (0.14) \\
SB        &       5.9 (0.9)      &         1.90 (0.14)   \\
\enddata
\end{deluxetable}

\newpage

\clearpage

\newpage
\begin{center}
\includegraphics[width=5cm,angle=180,keepaspectratio=true]{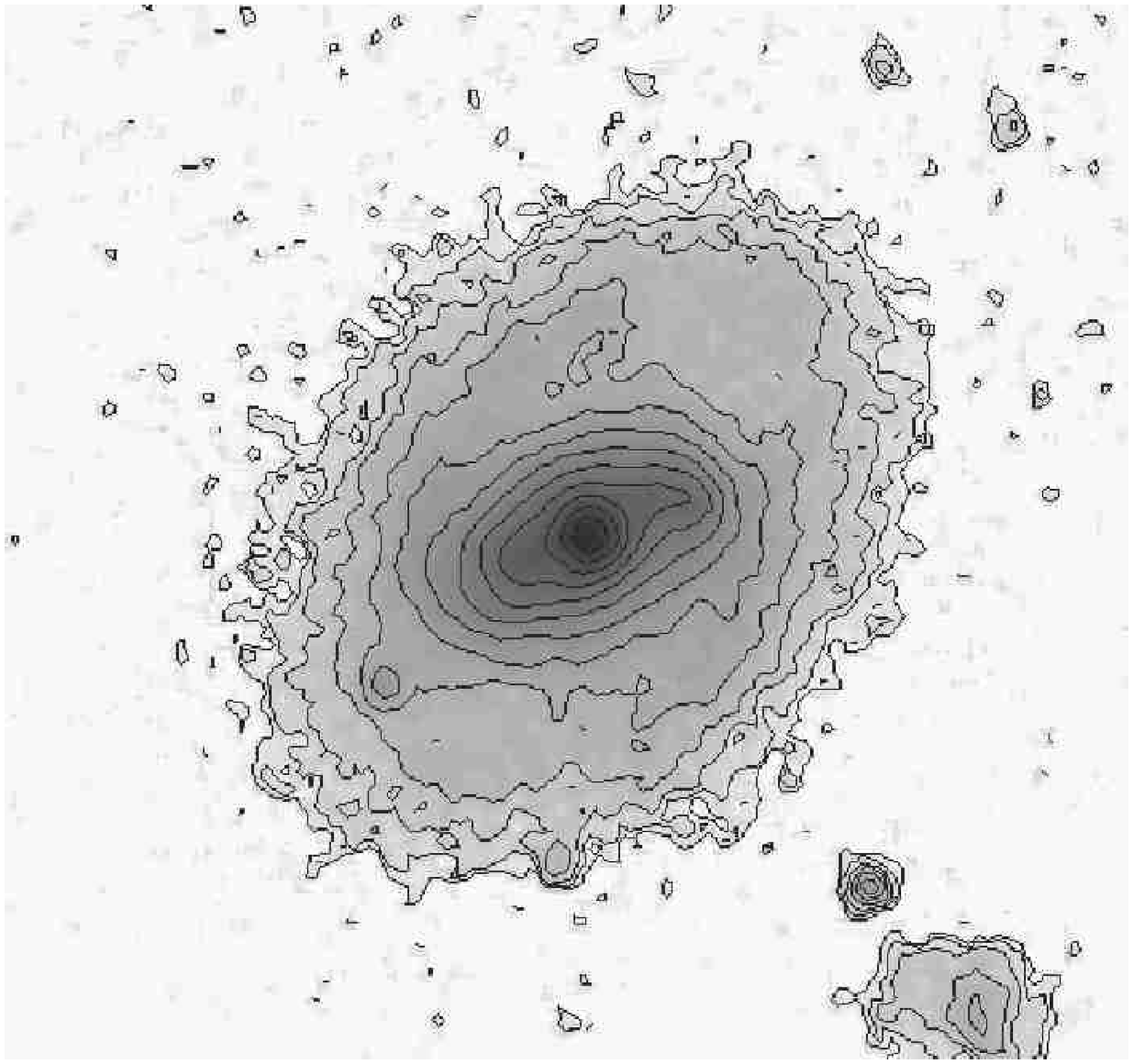}
\includegraphics[width=5cm,angle=180,keepaspectratio=true]{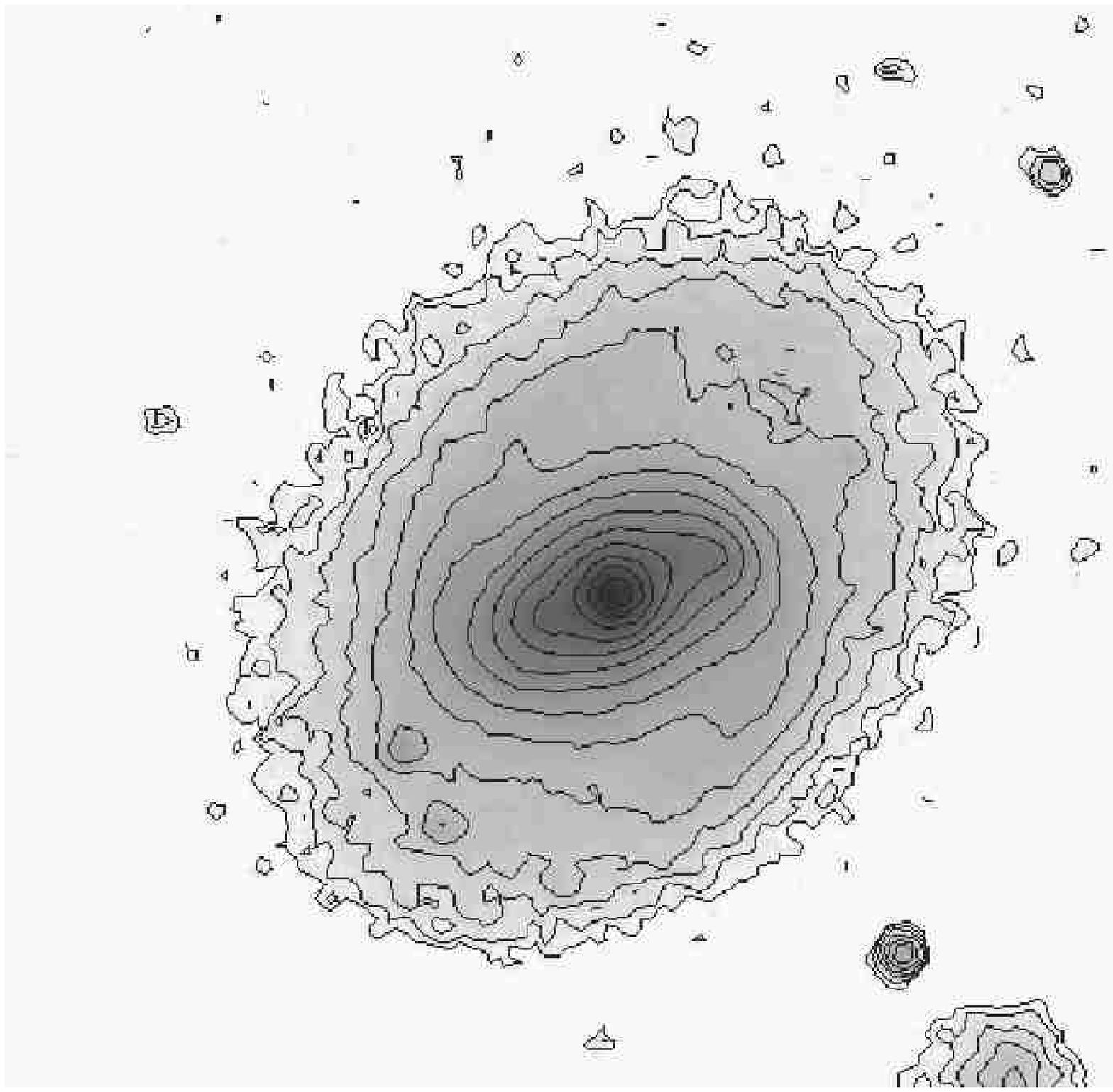} \\
\includegraphics[width=5cm,angle=180,keepaspectratio=true]{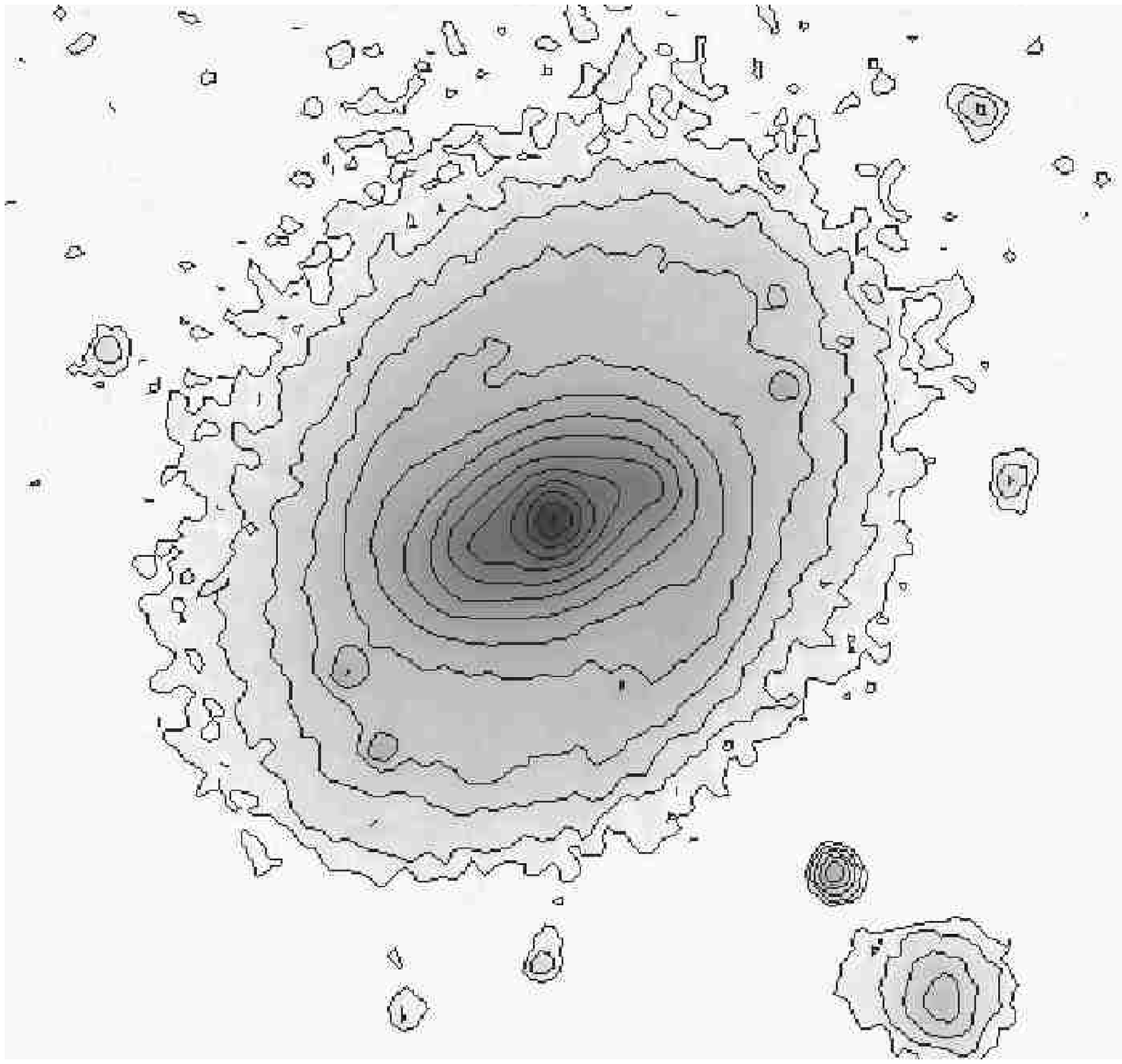}
\includegraphics[width=5cm,angle=180,keepaspectratio=true]{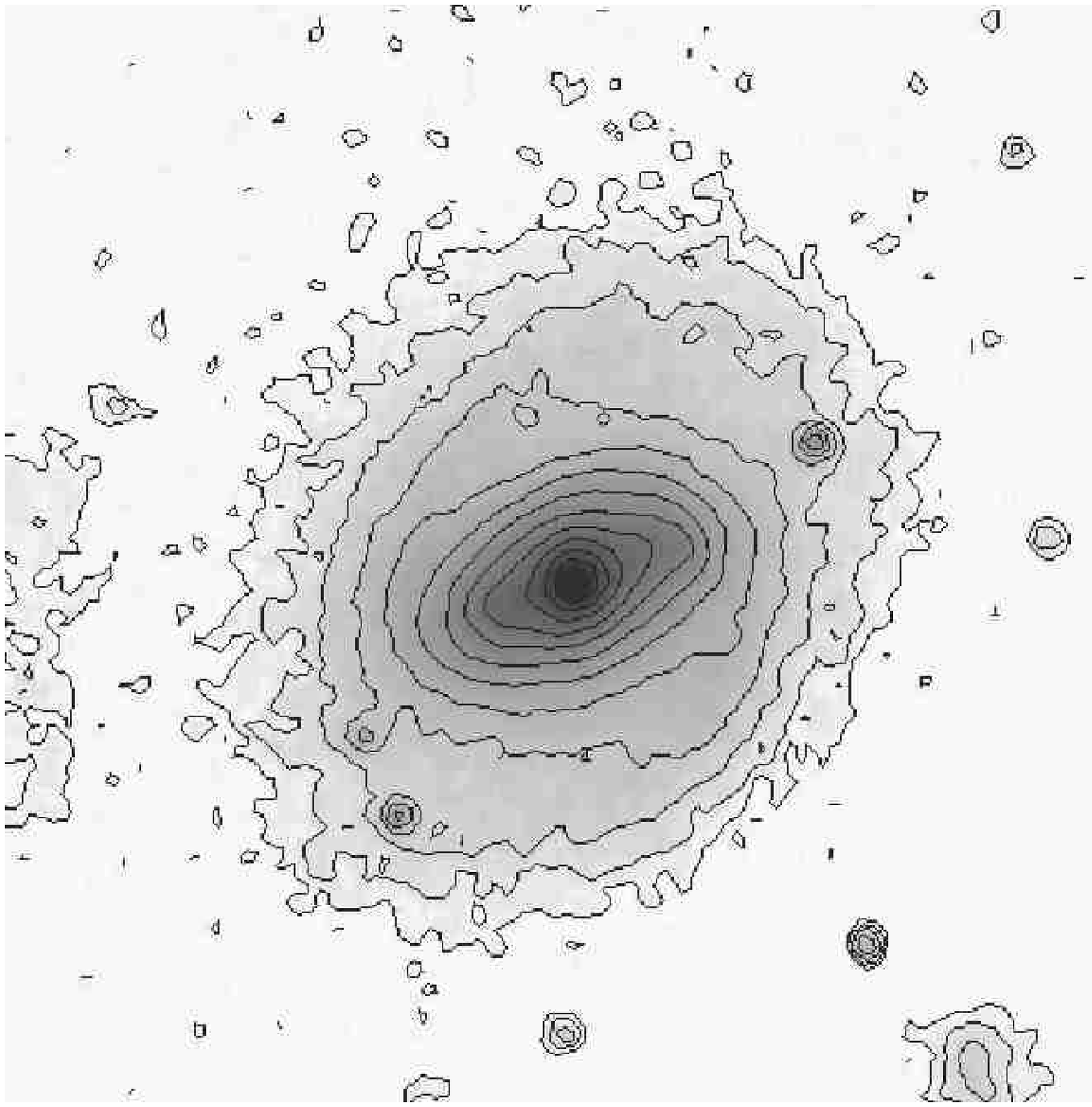} \\
\vskip 1.25cm \includegraphics[width=10cm,keepaspectratio=true]{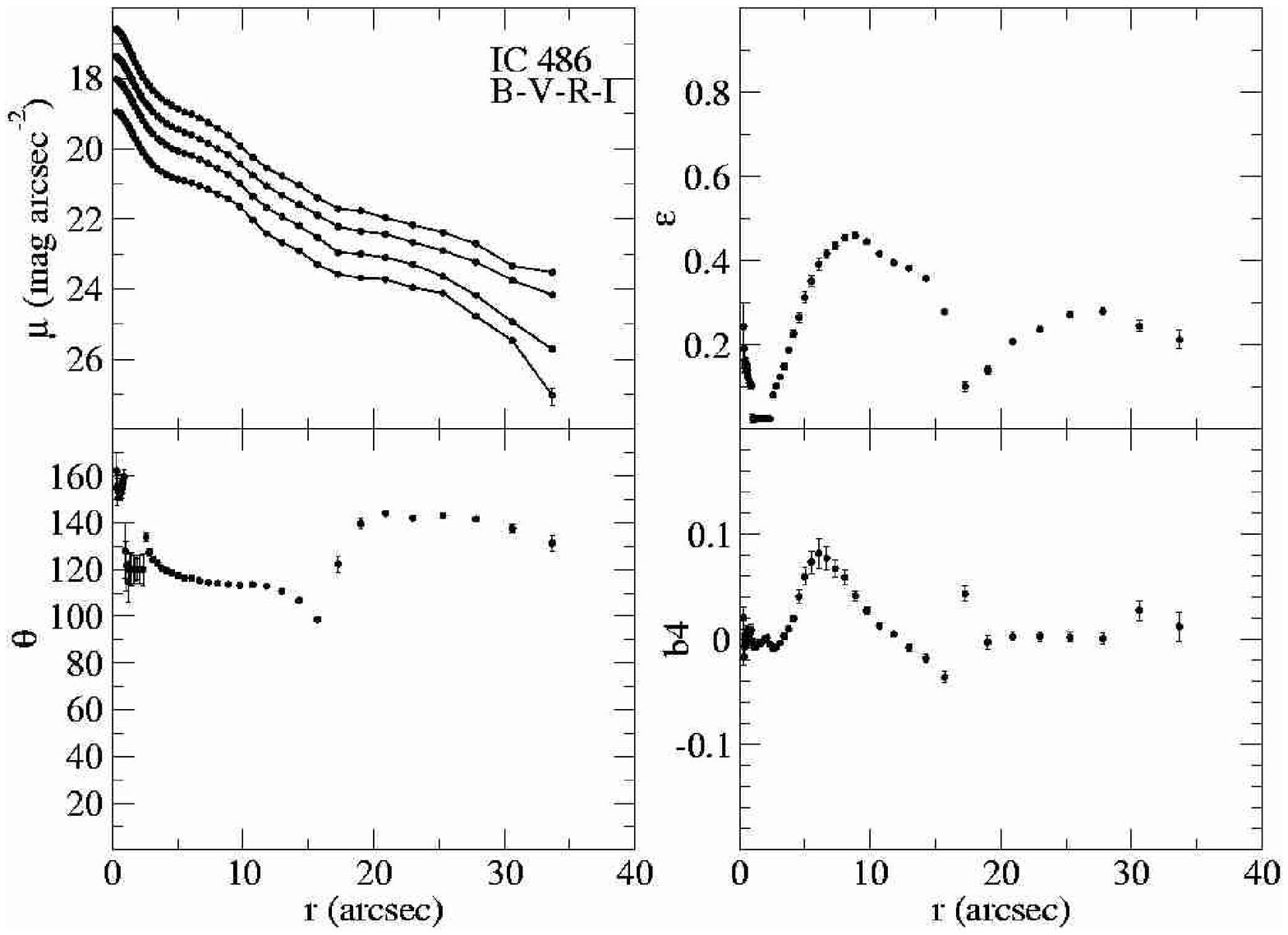}
\end{center}

\newpage
\begin{figure}
\begin{center}
\includegraphics[width=10cm,keepaspectratio=true]{f2.eps}
\end{center}
\end{figure}

\newpage
\begin{figure}
\begin{center}
\vskip 0.5cm \includegraphics[width=7cm,keepaspectratio=true]{f3a.eps}
\hskip 1cm \includegraphics[width=7cm,keepaspectratio=true]{f3b.eps} \\
\vskip 1.25cm \includegraphics[width=7cm,keepaspectratio=true]{f3c.eps}
\hskip 1cm \includegraphics[width=7cm,keepaspectratio=true]{f3d.eps}
\end{center}
\end{figure}

\begin{figure}
\includegraphics[width=4cm,height=4cm,keepaspectratio=false]{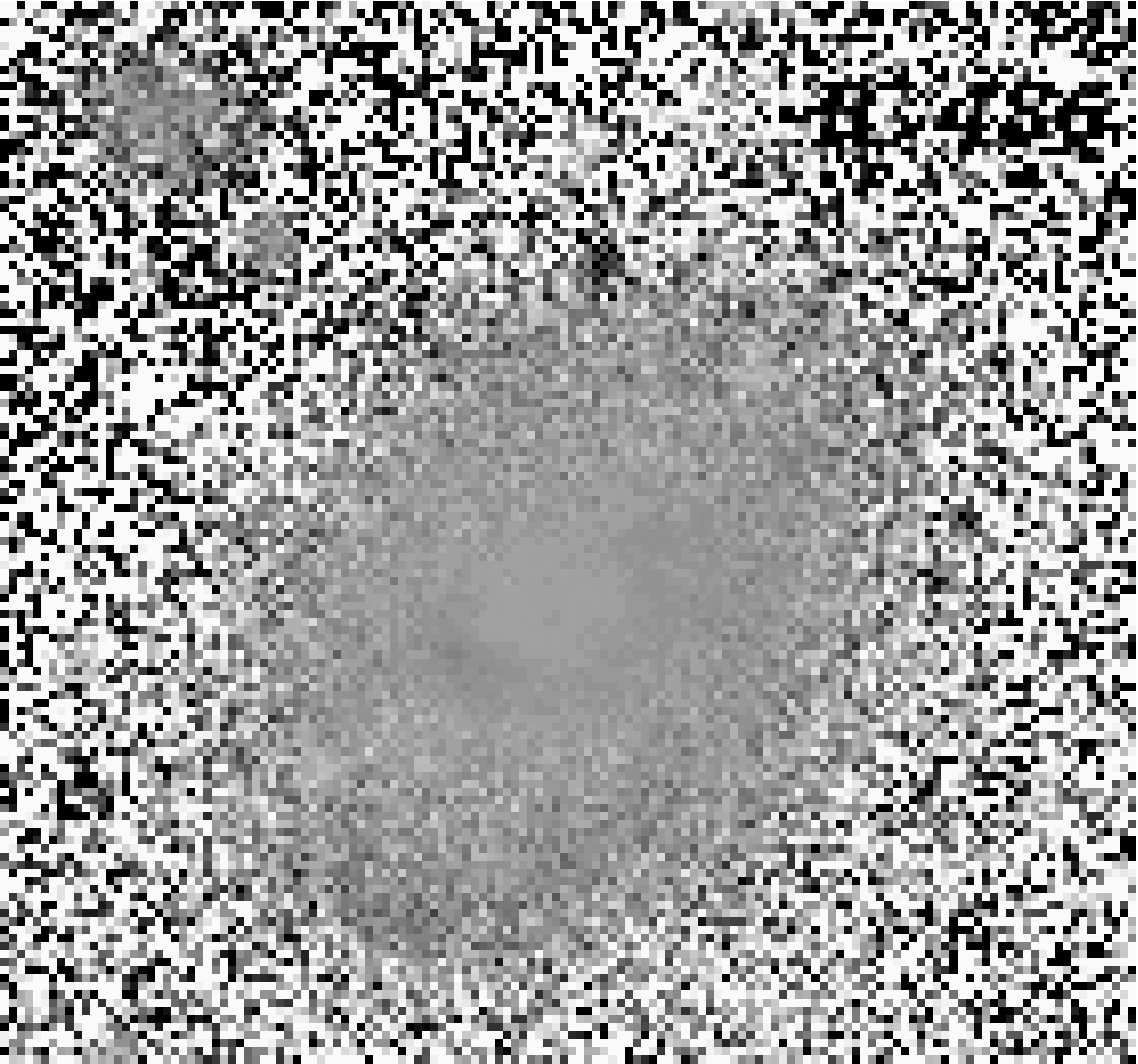}
\includegraphics[width=4cm,height=4cm,keepaspectratio=false]{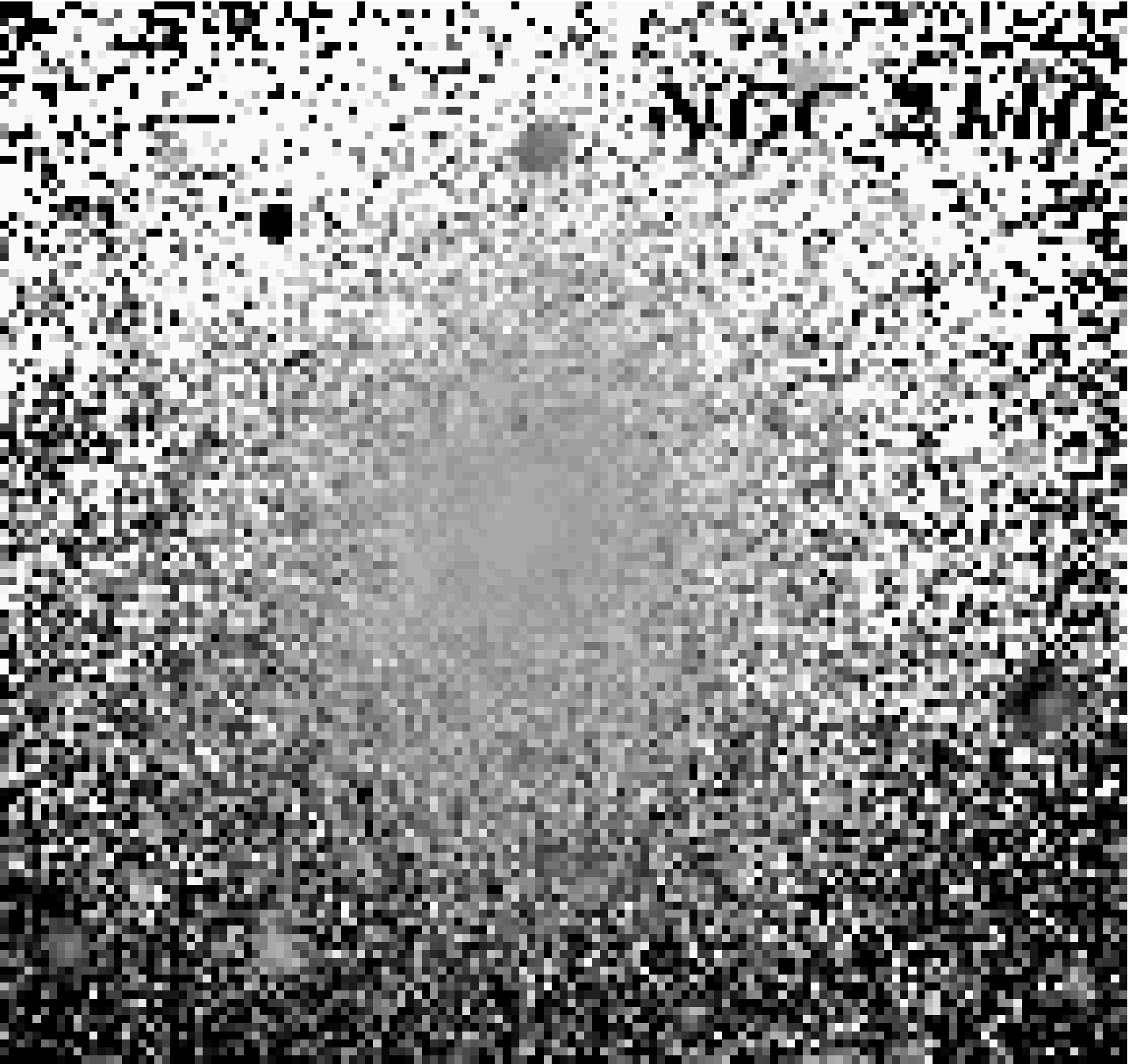}
\includegraphics[width=4cm,height=4cm,keepaspectratio=false]{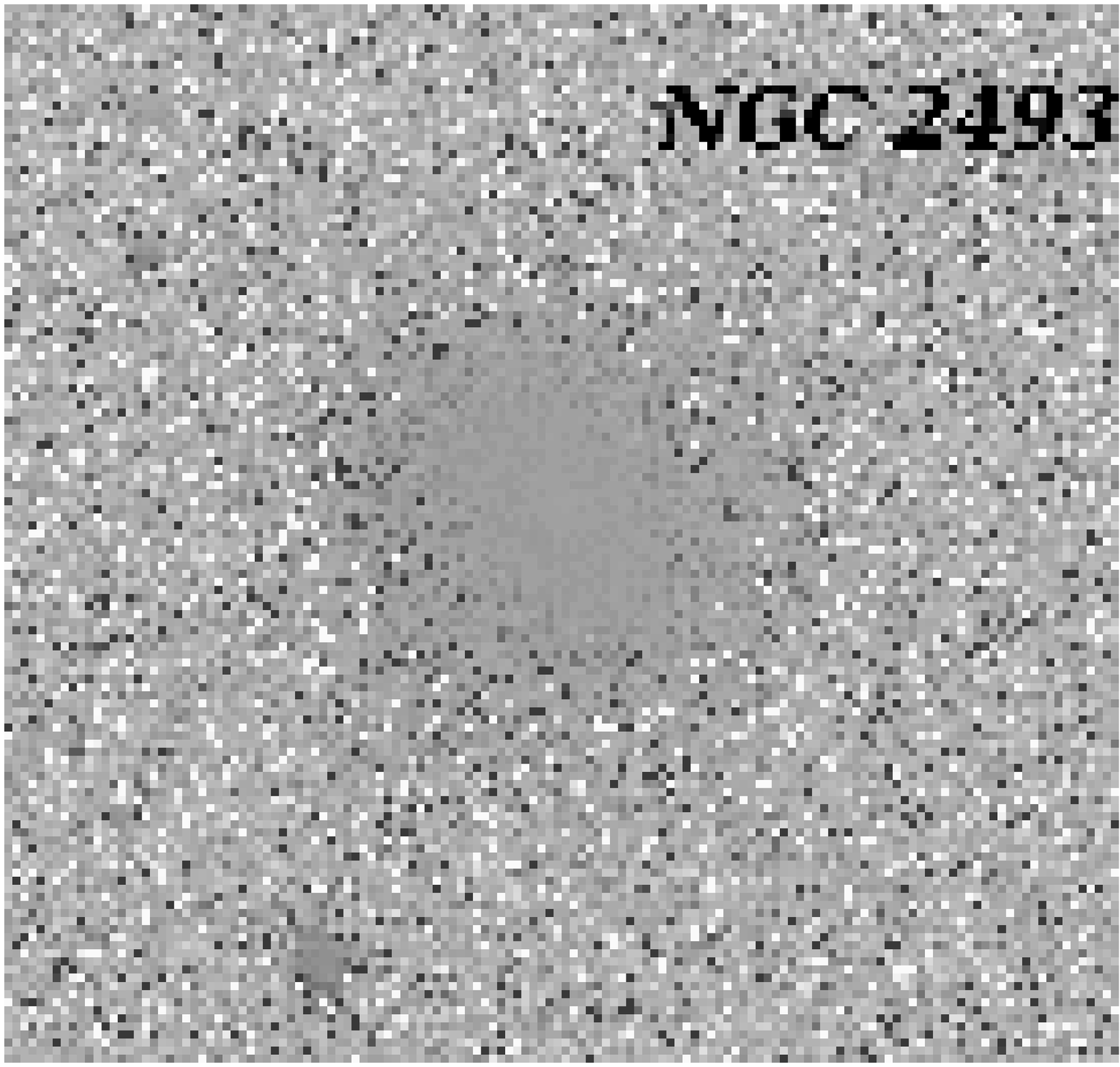}
\includegraphics[width=4cm,height=4cm,keepaspectratio=false]{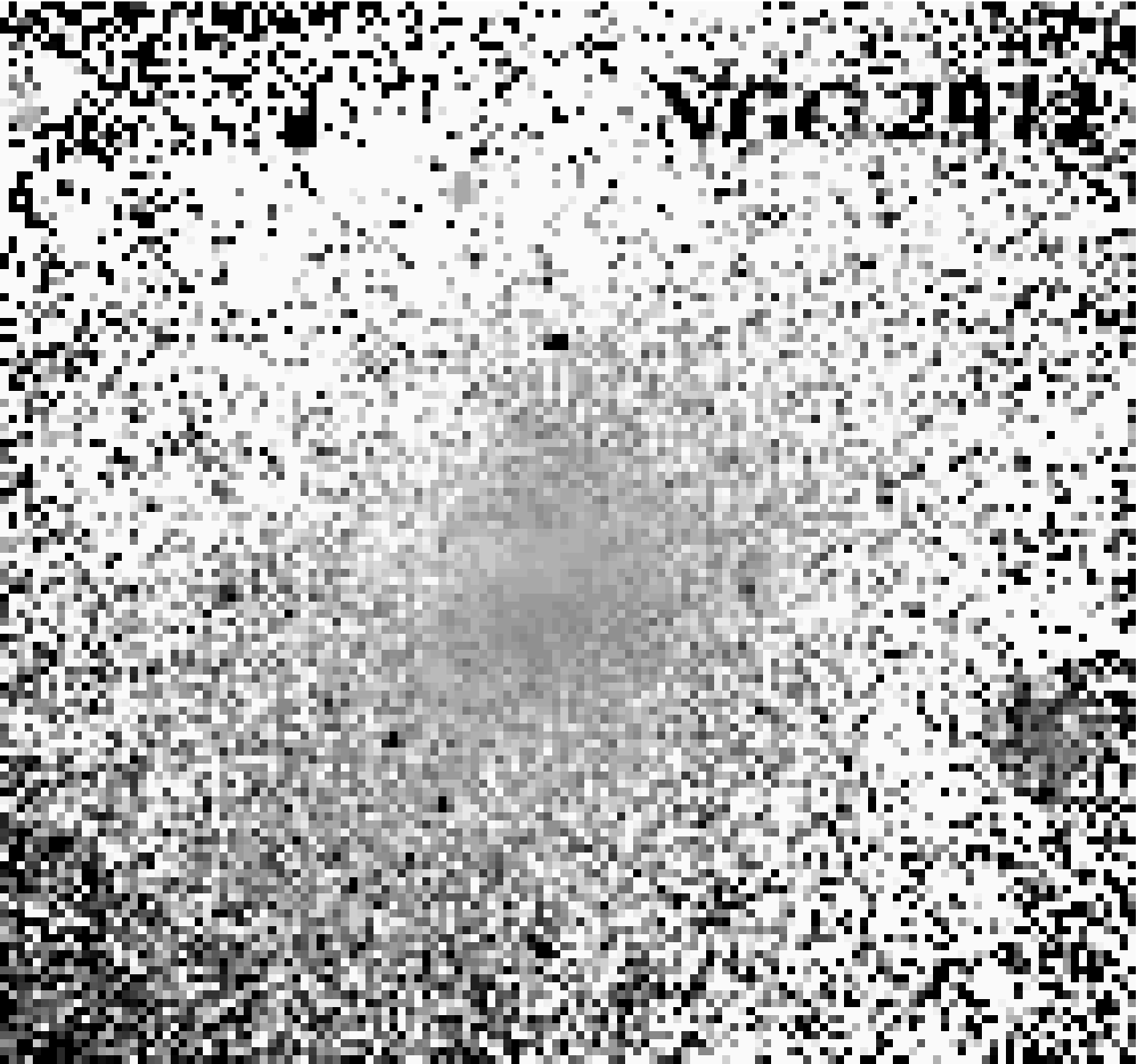}\\
\includegraphics[width=4cm,height=4cm,keepaspectratio=false]{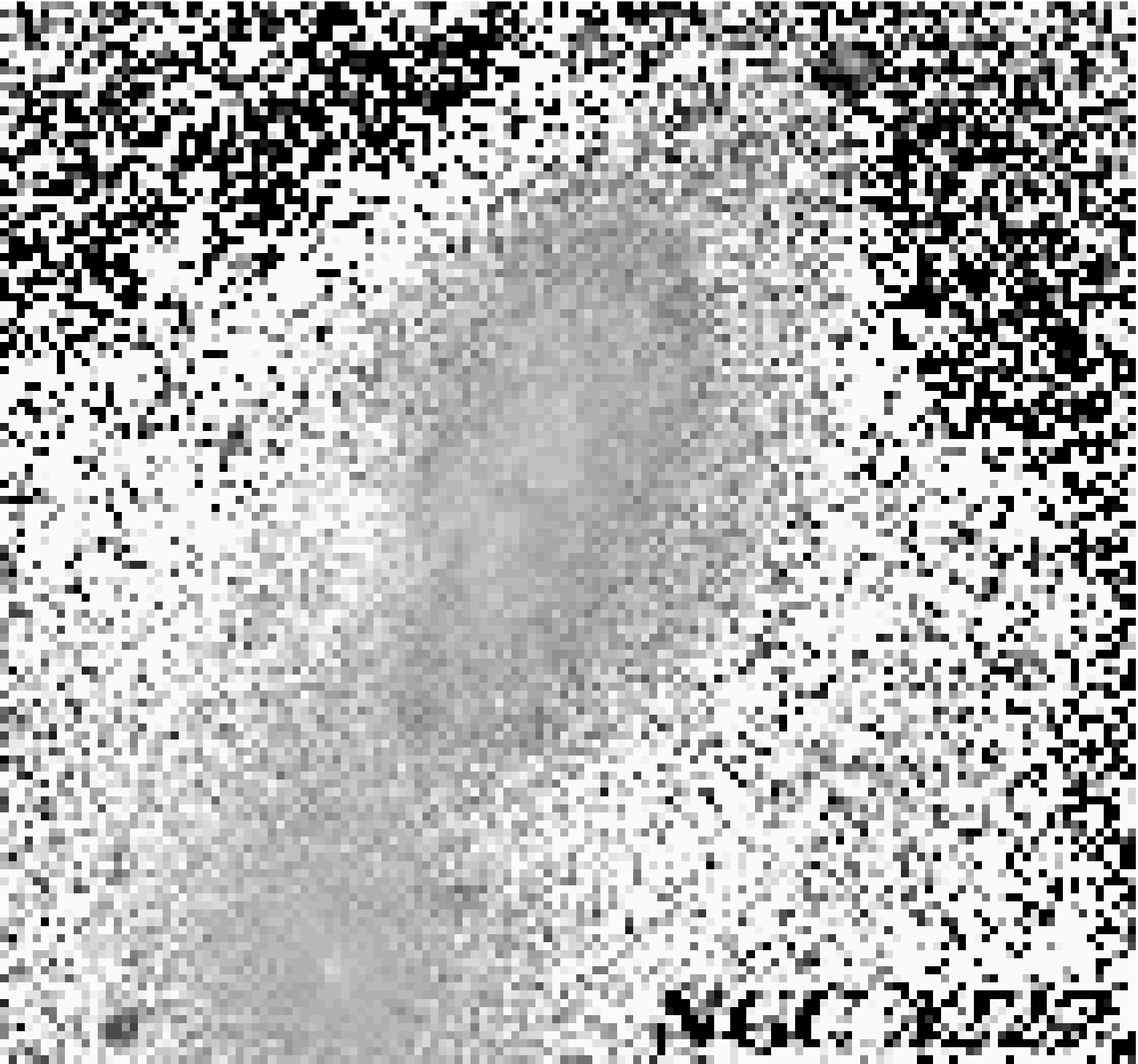}
\includegraphics[width=4cm,height=4cm,keepaspectratio=false]{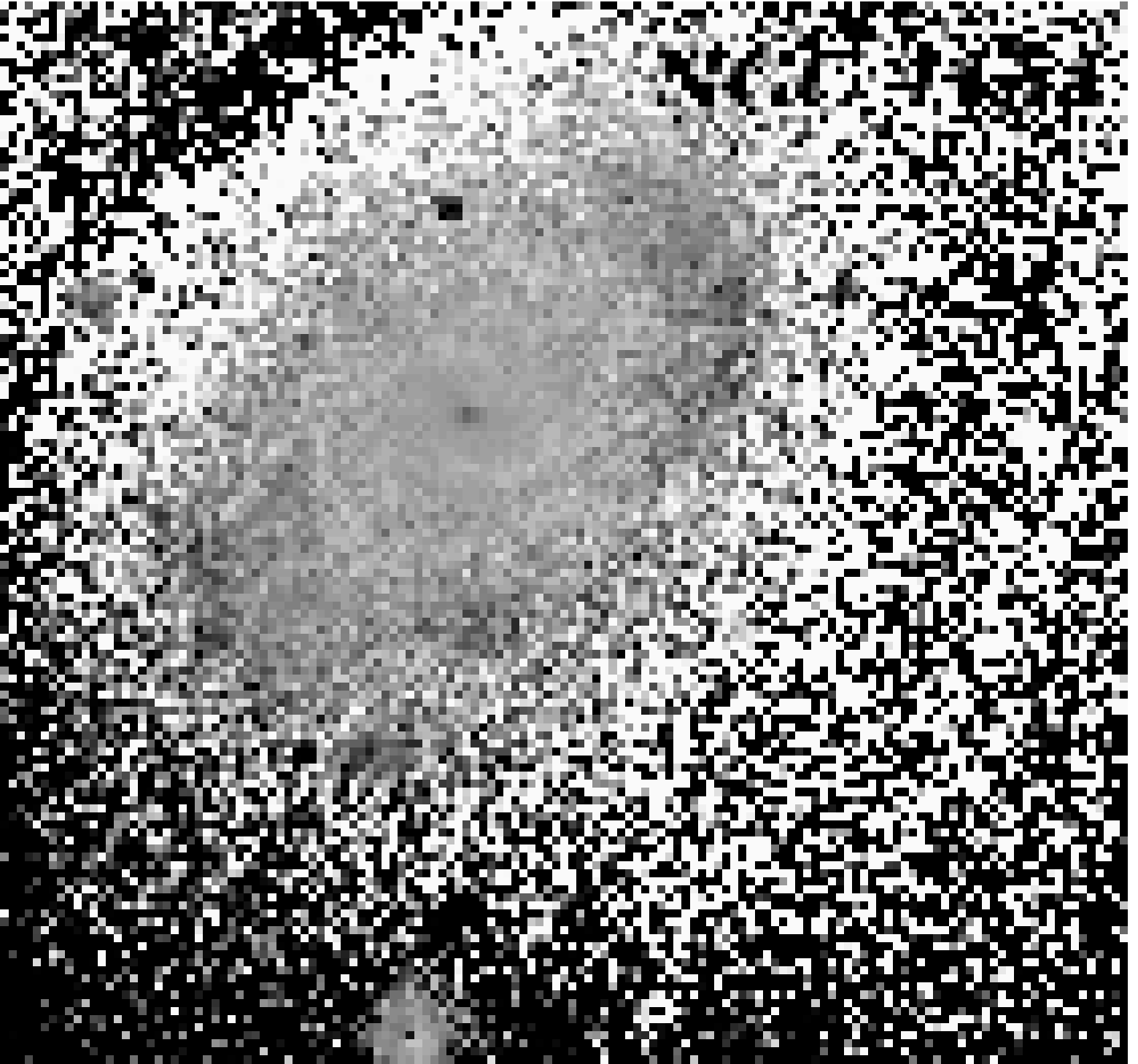}
\includegraphics[width=4cm,height=4cm,keepaspectratio=false]{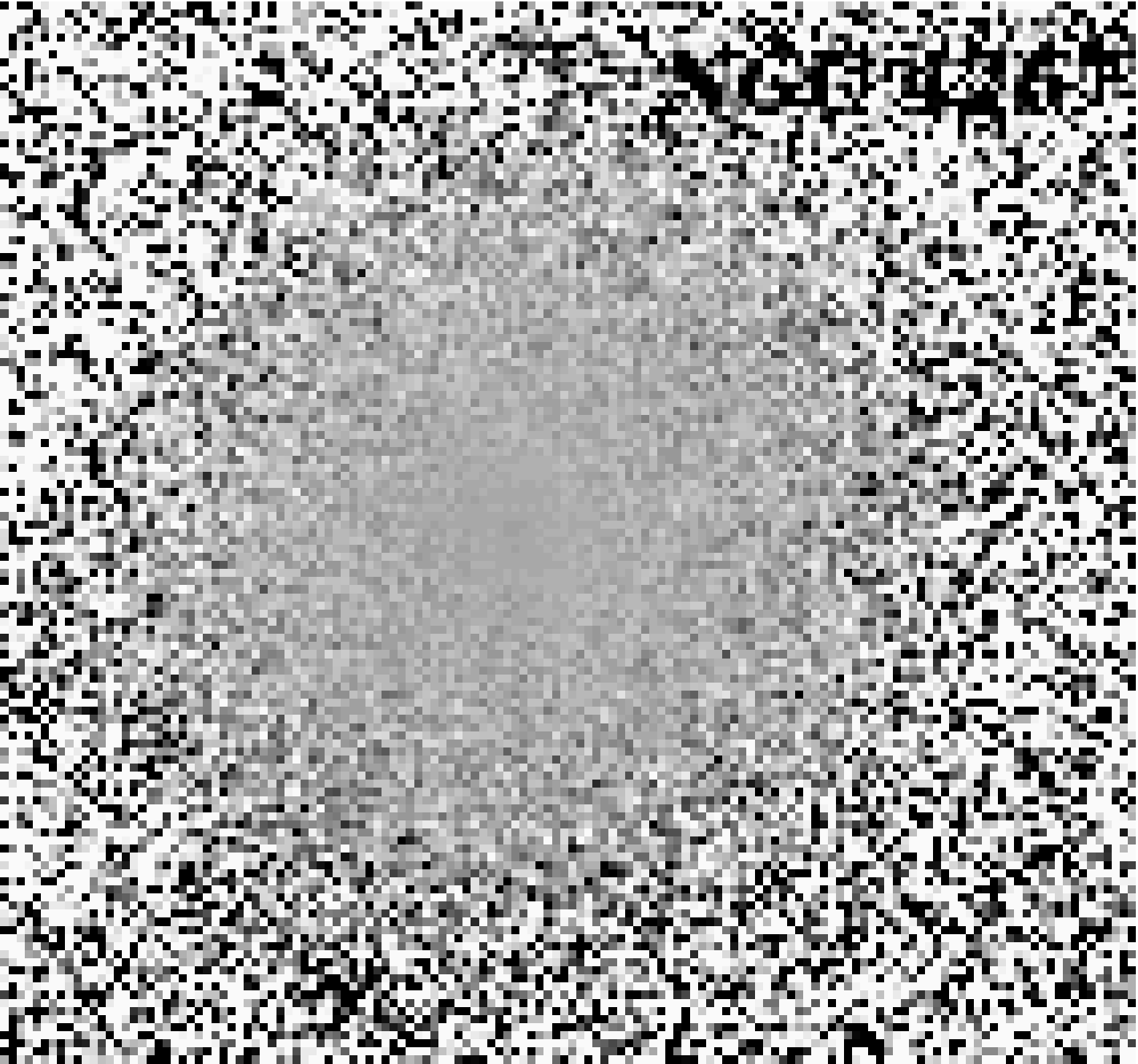}
\includegraphics[width=4cm,height=4cm,keepaspectratio=false]{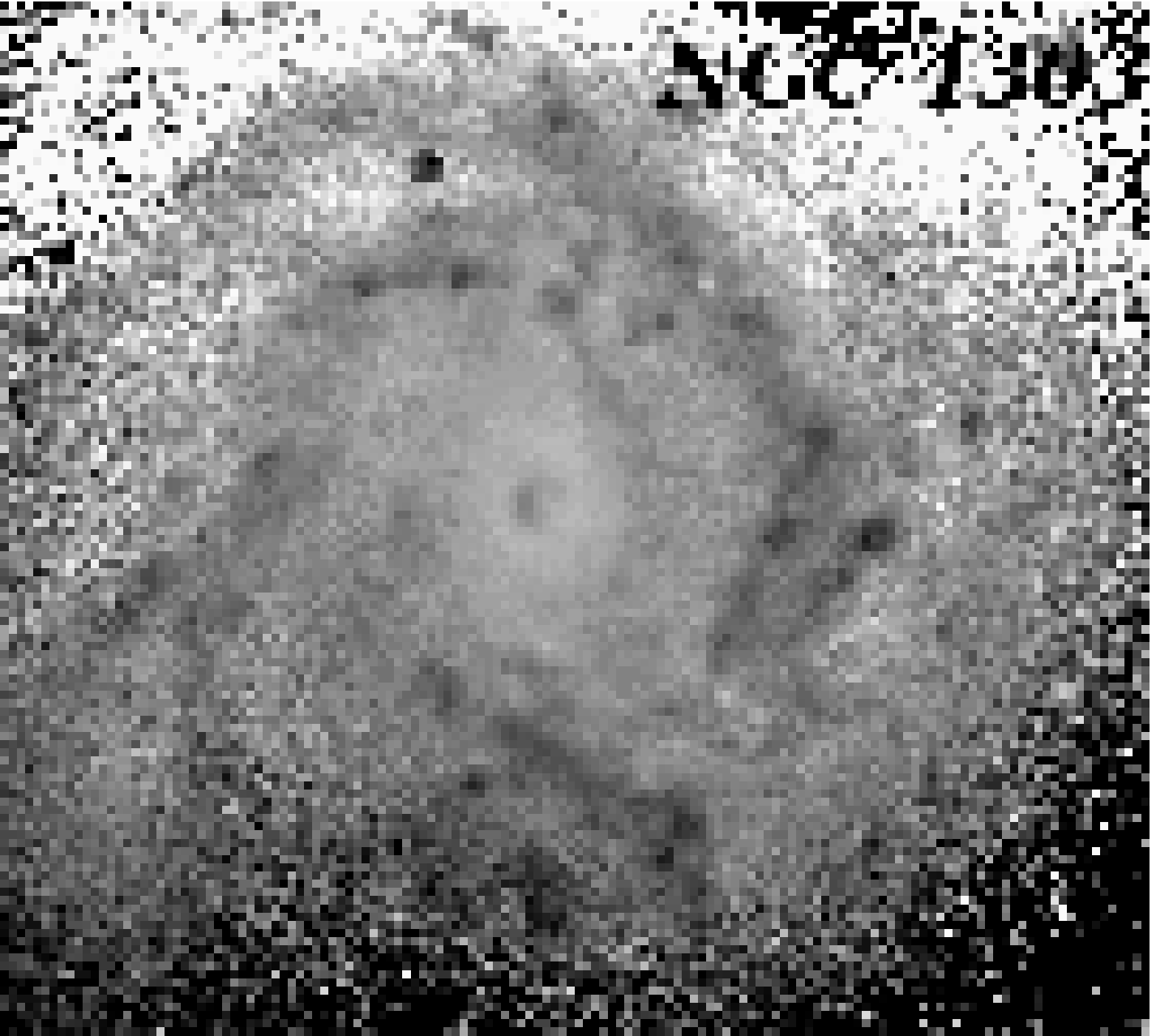}\\
\includegraphics[width=4cm,height=4cm,keepaspectratio=false]{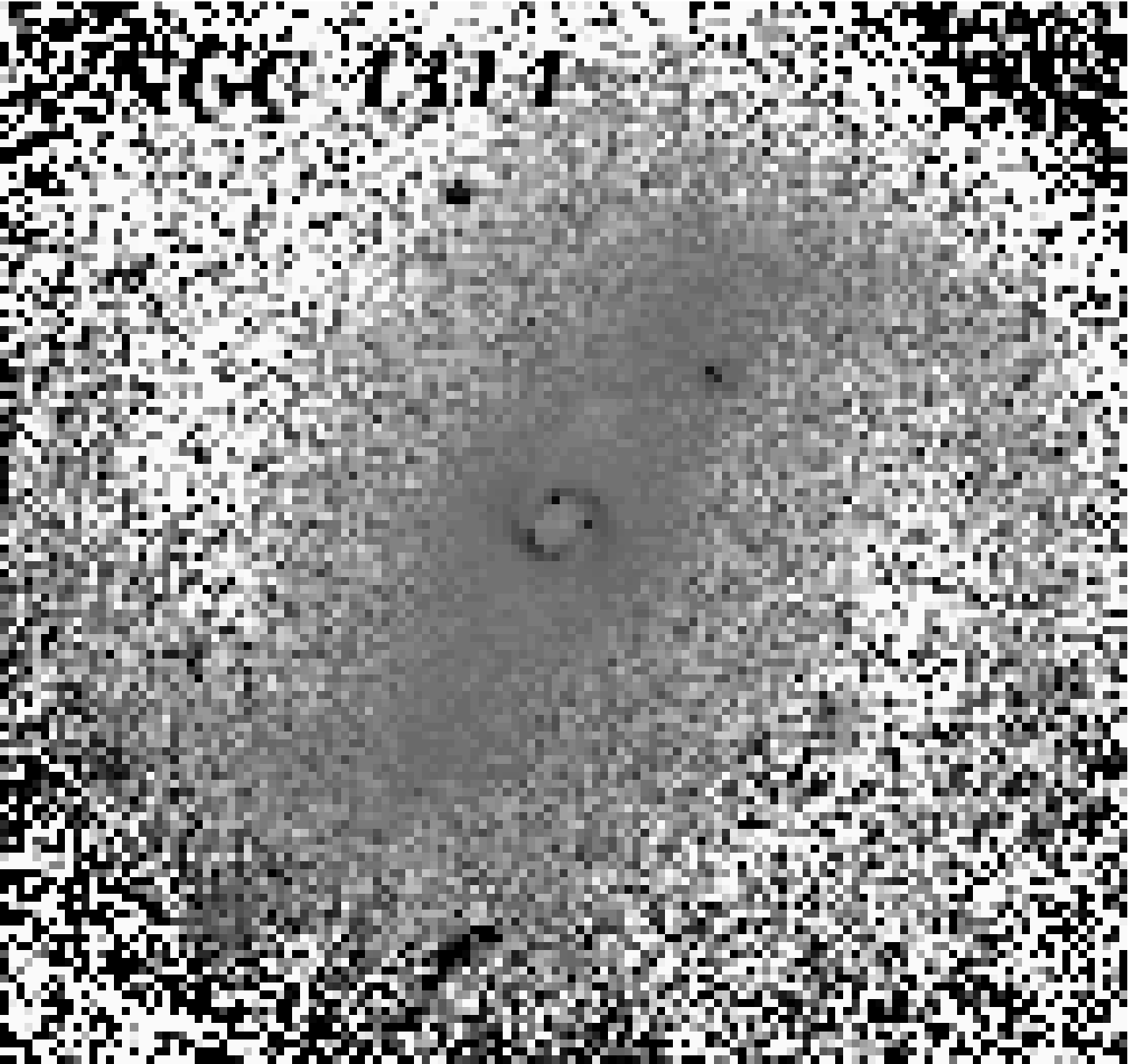}
\includegraphics[width=4cm,height=4cm,keepaspectratio=false]{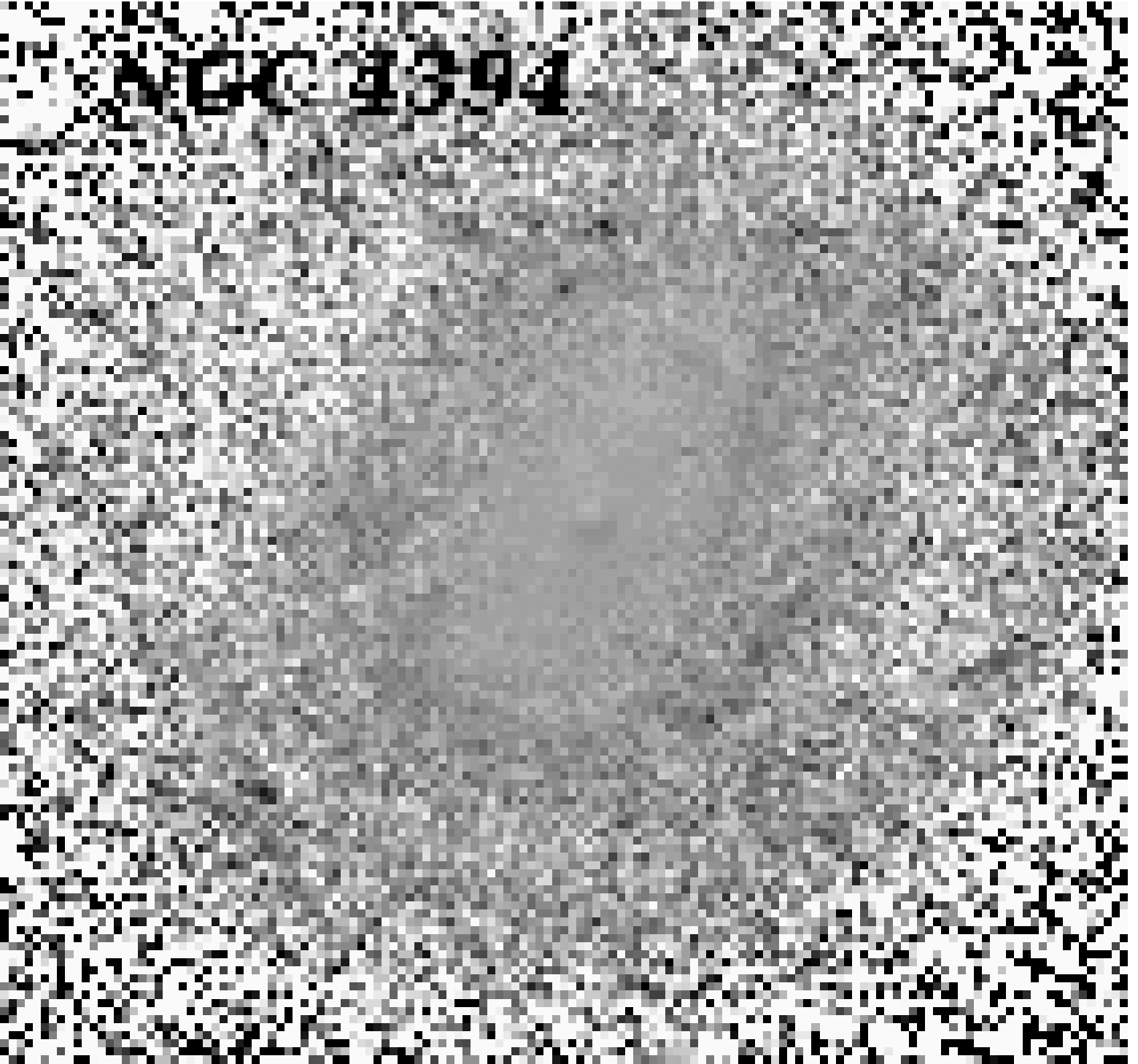}
\includegraphics[width=4cm,height=4cm,keepaspectratio=false]{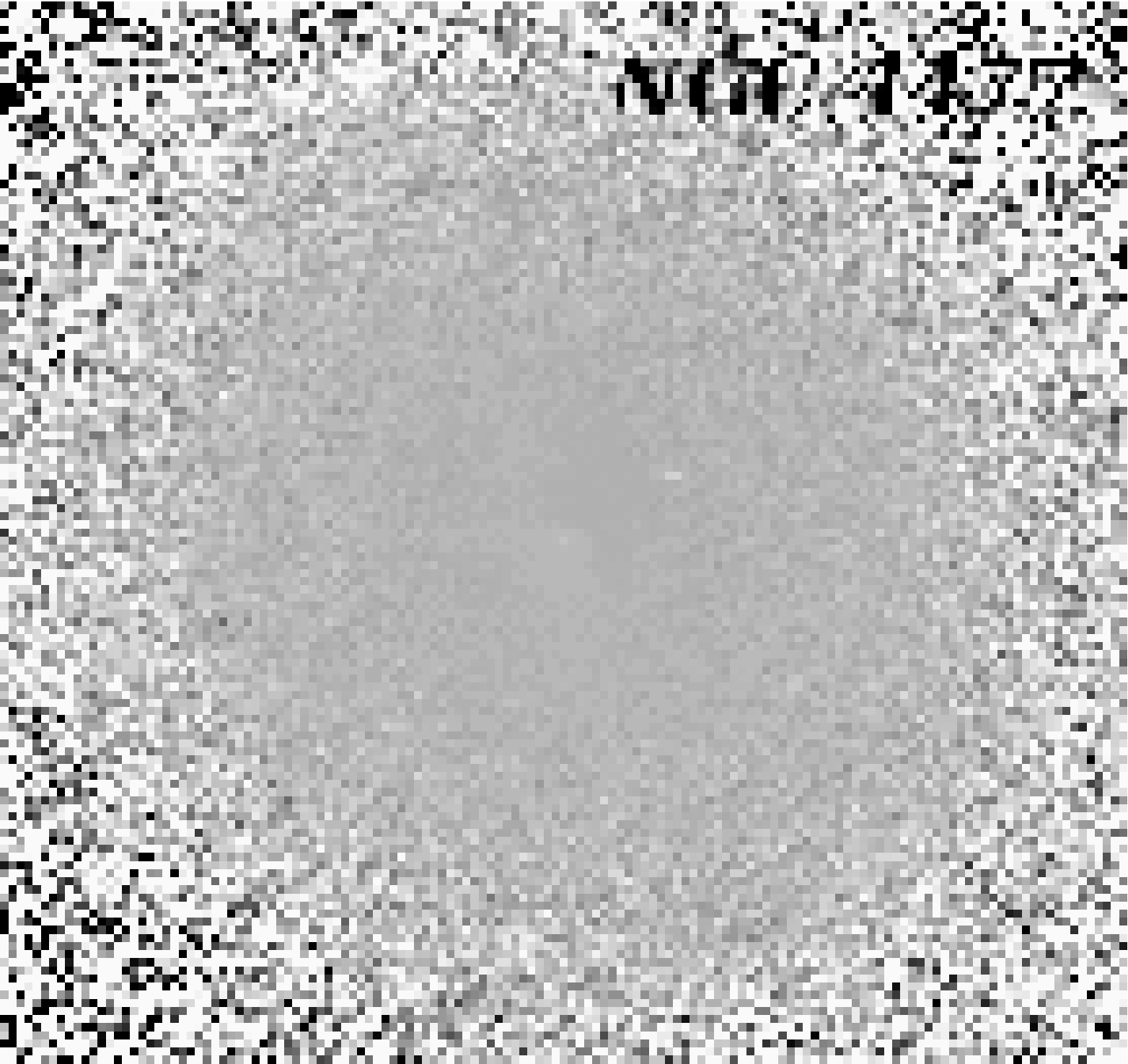}
\includegraphics[width=4cm,height=4cm,keepaspectratio=false]{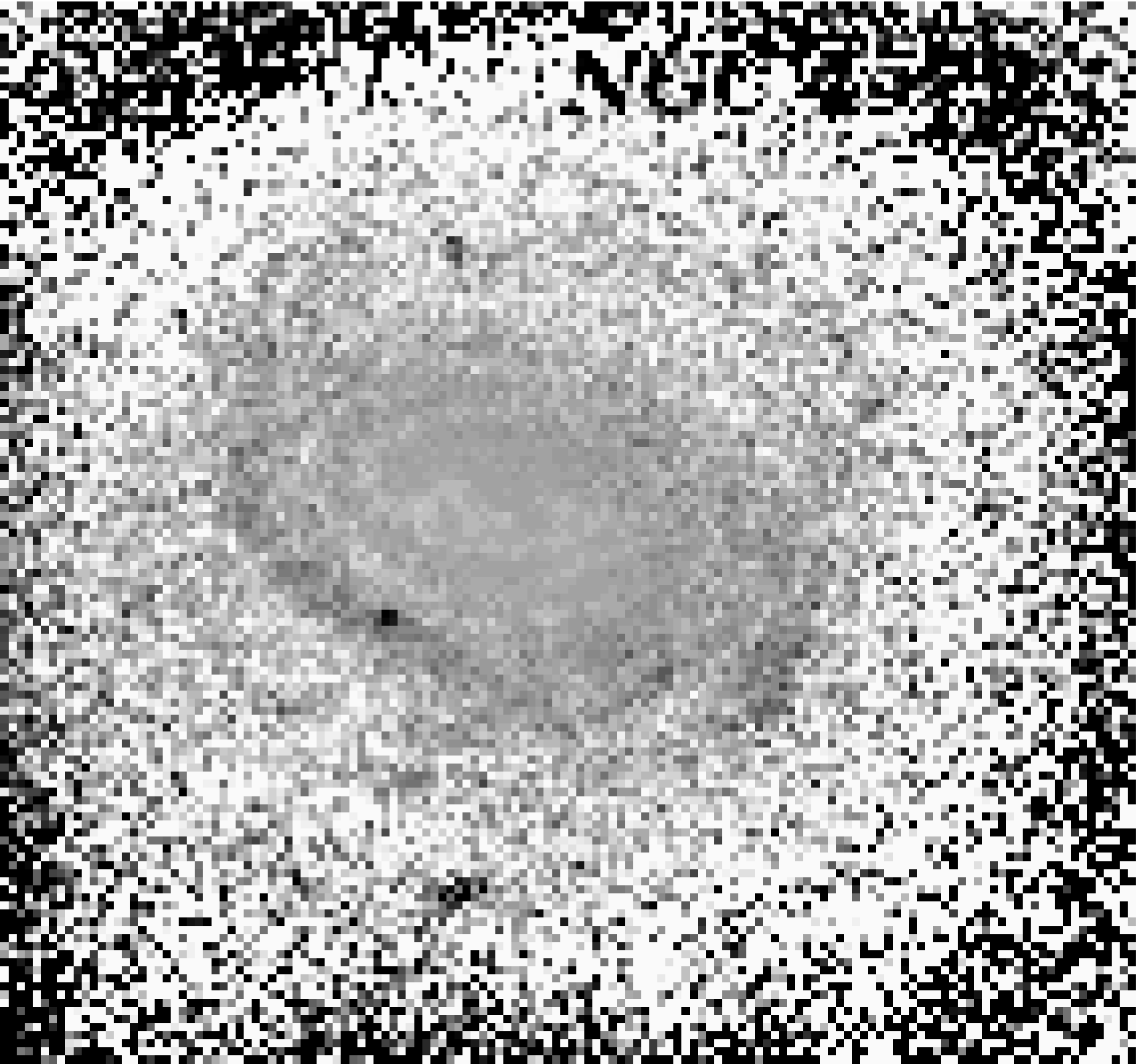}\\
\includegraphics[width=4cm,height=4cm,keepaspectratio=false]{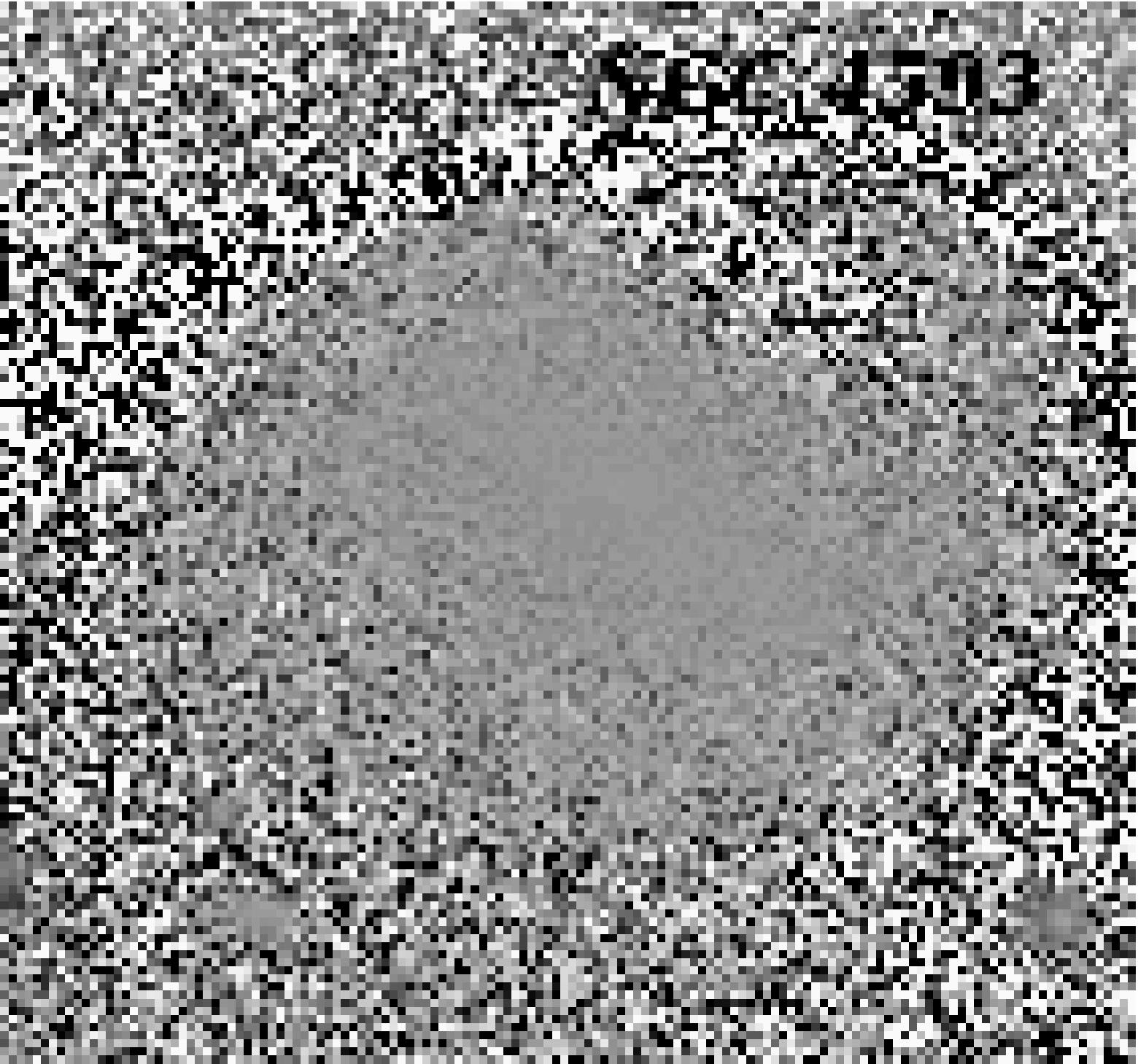}
\includegraphics[width=4cm,height=4cm,keepaspectratio=false]{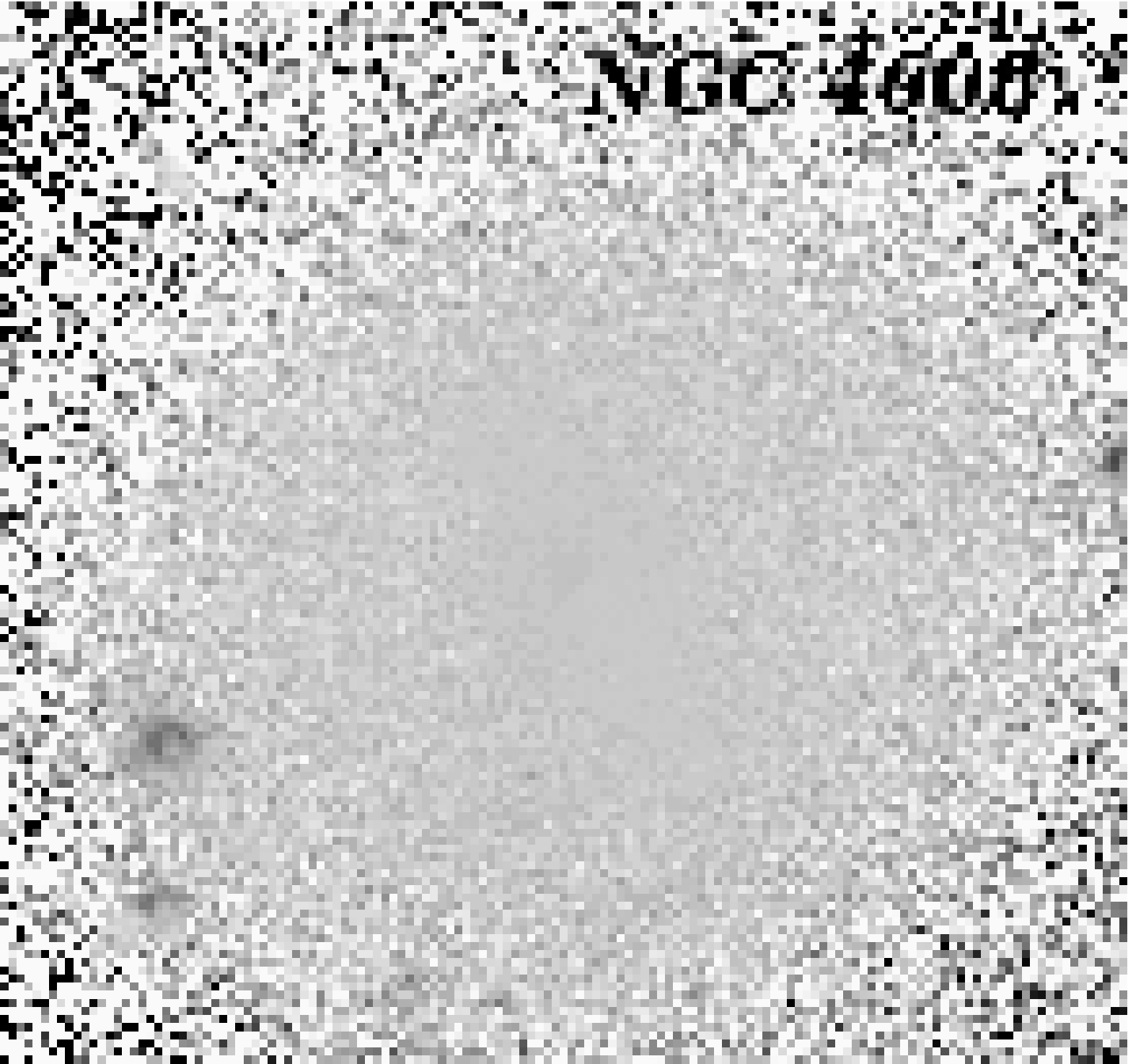}
\includegraphics[width=4cm,height=4cm,keepaspectratio=false]{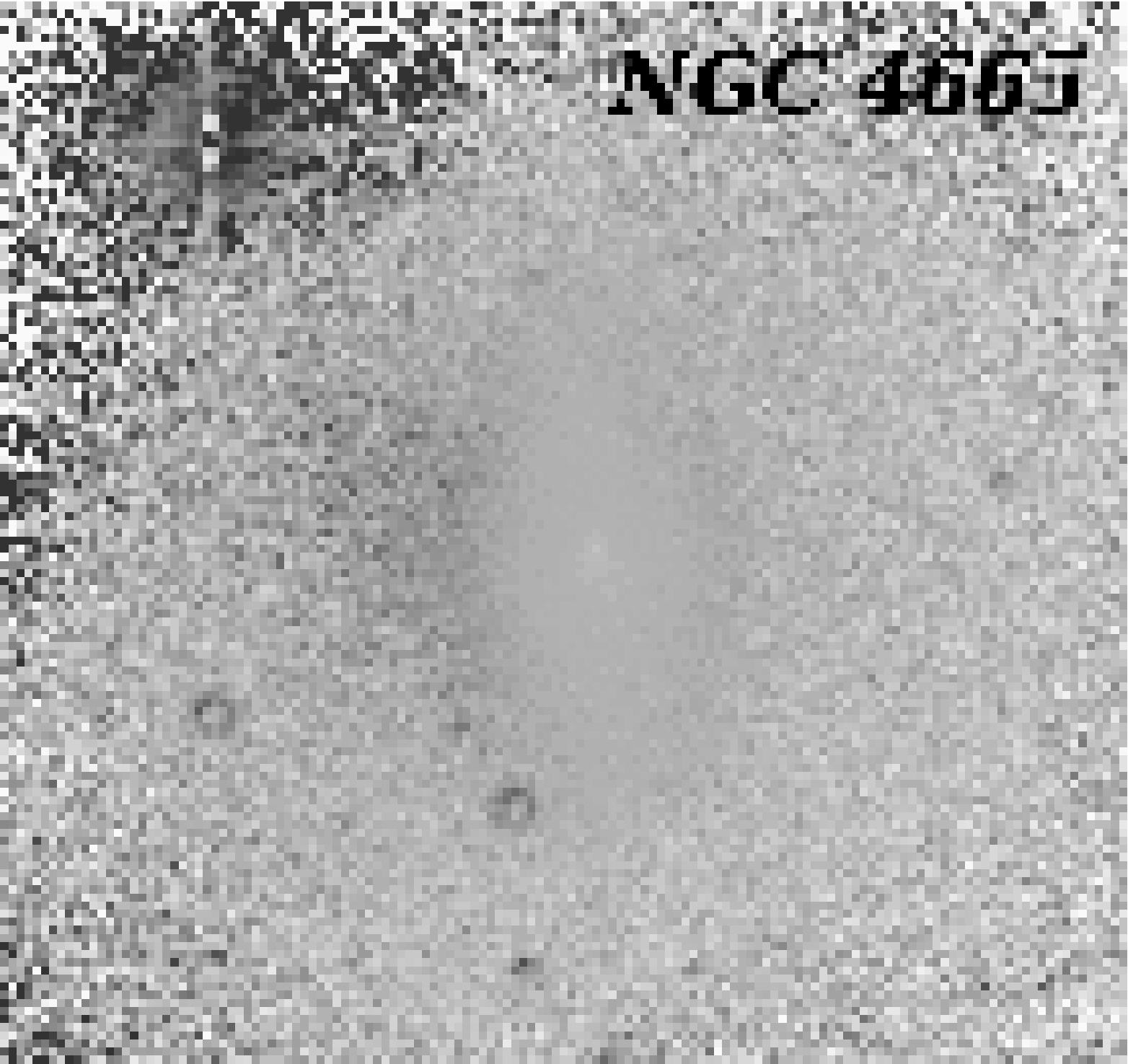}
\includegraphics[width=4cm,height=4cm,keepaspectratio=false]{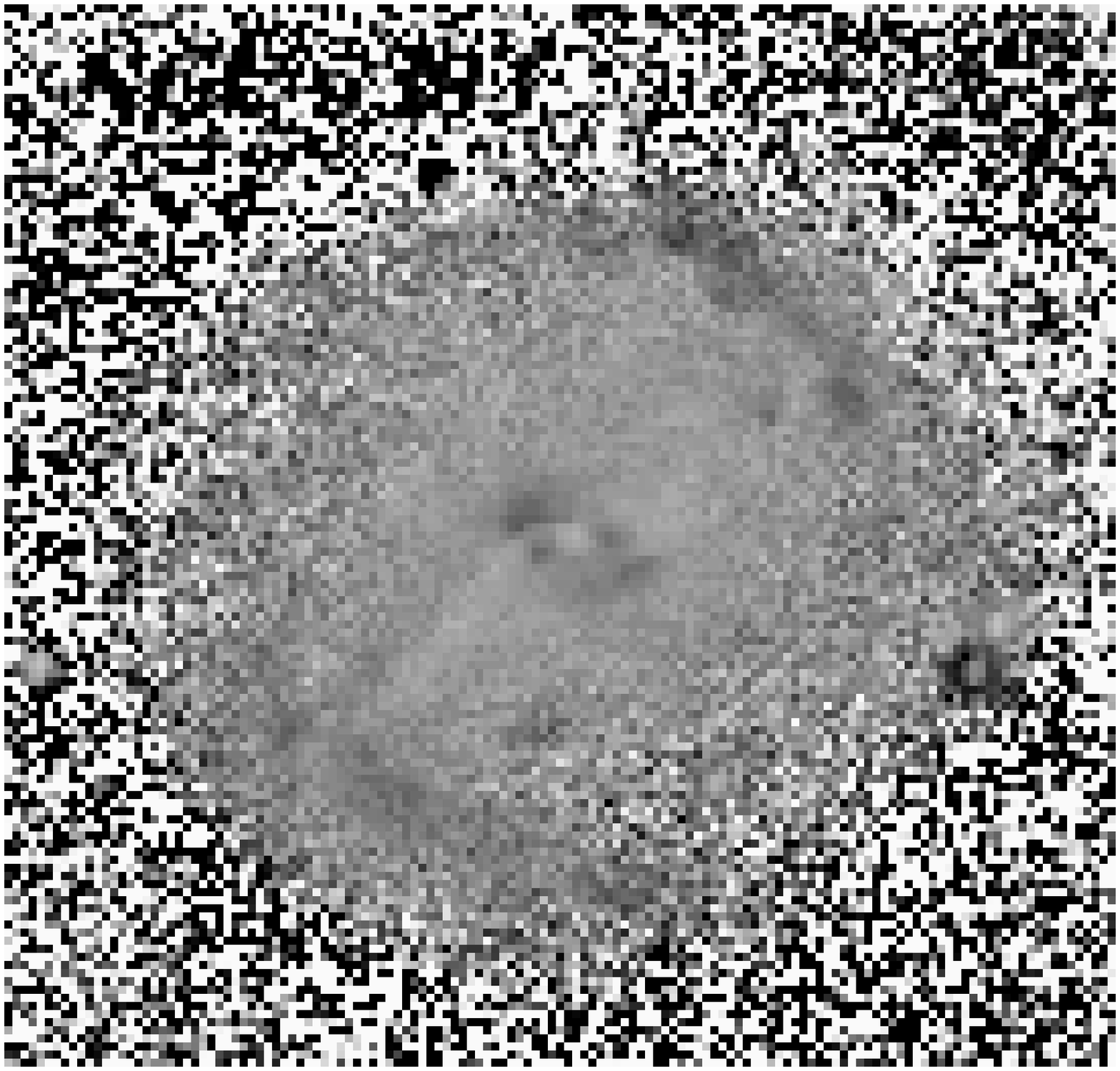}\\
\includegraphics[width=4cm,height=4cm,keepaspectratio=false]{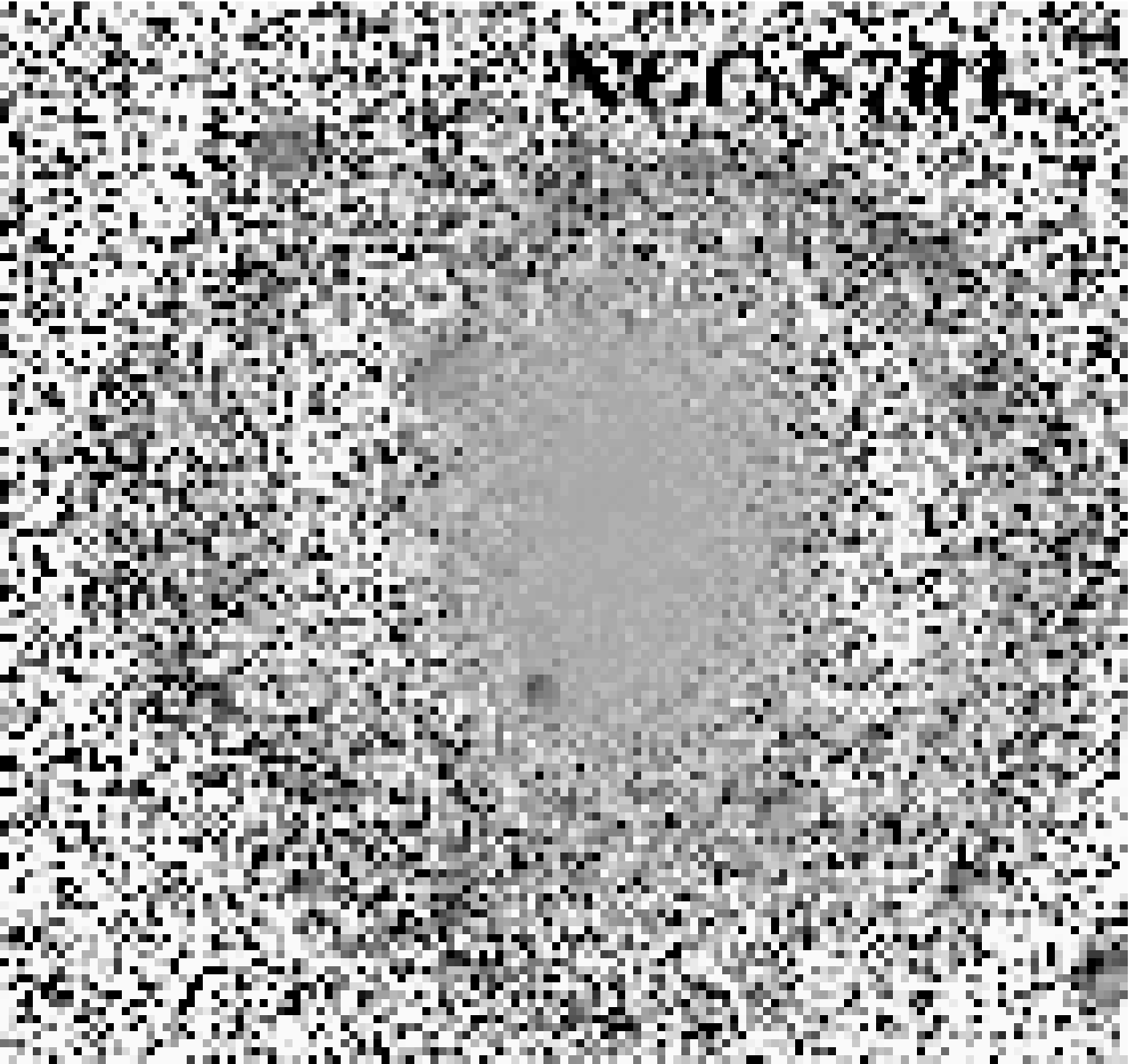}
\includegraphics[width=4cm,height=4cm,keepaspectratio=false]{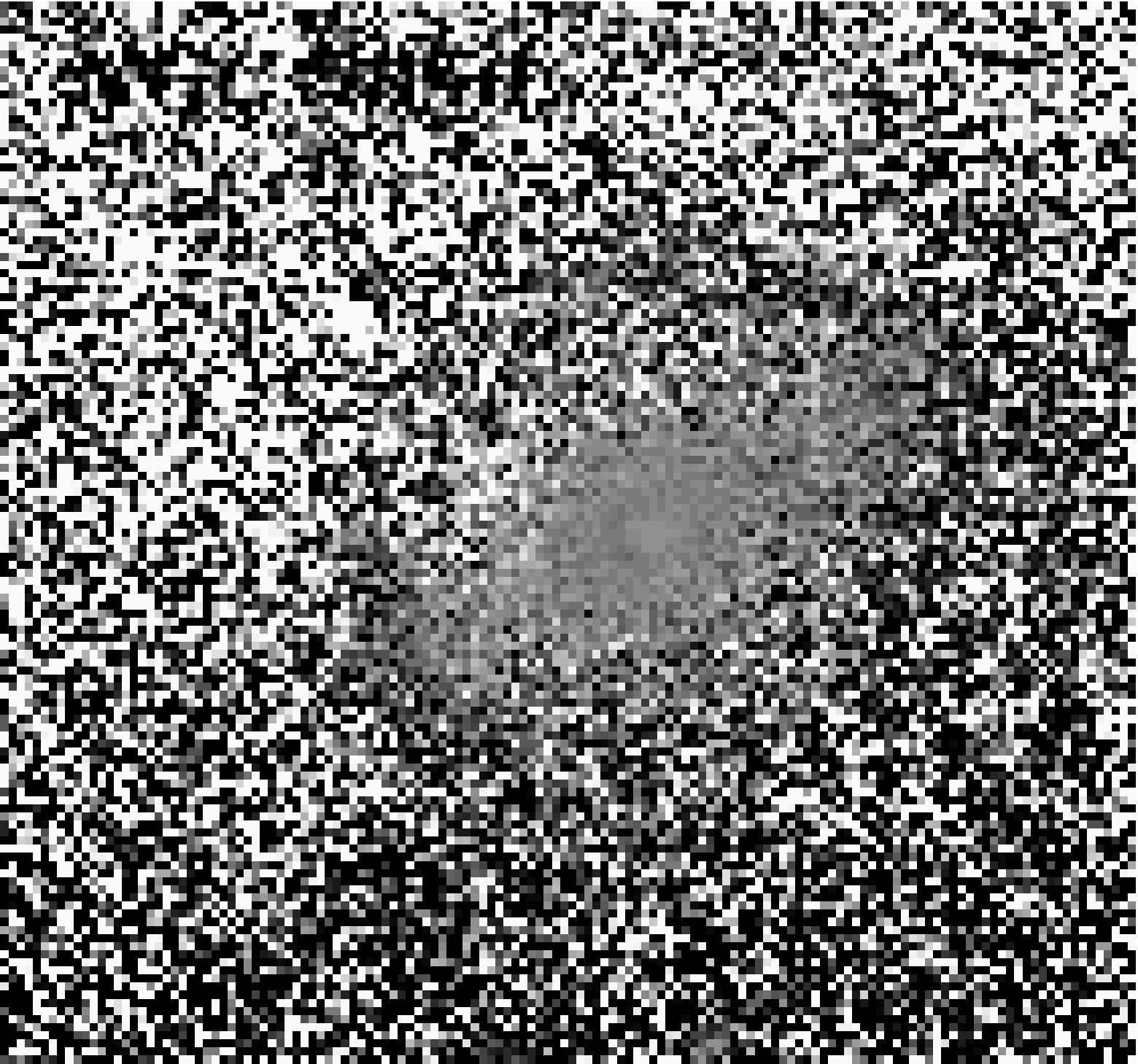}
\includegraphics[width=4cm,height=4cm,keepaspectratio=false]{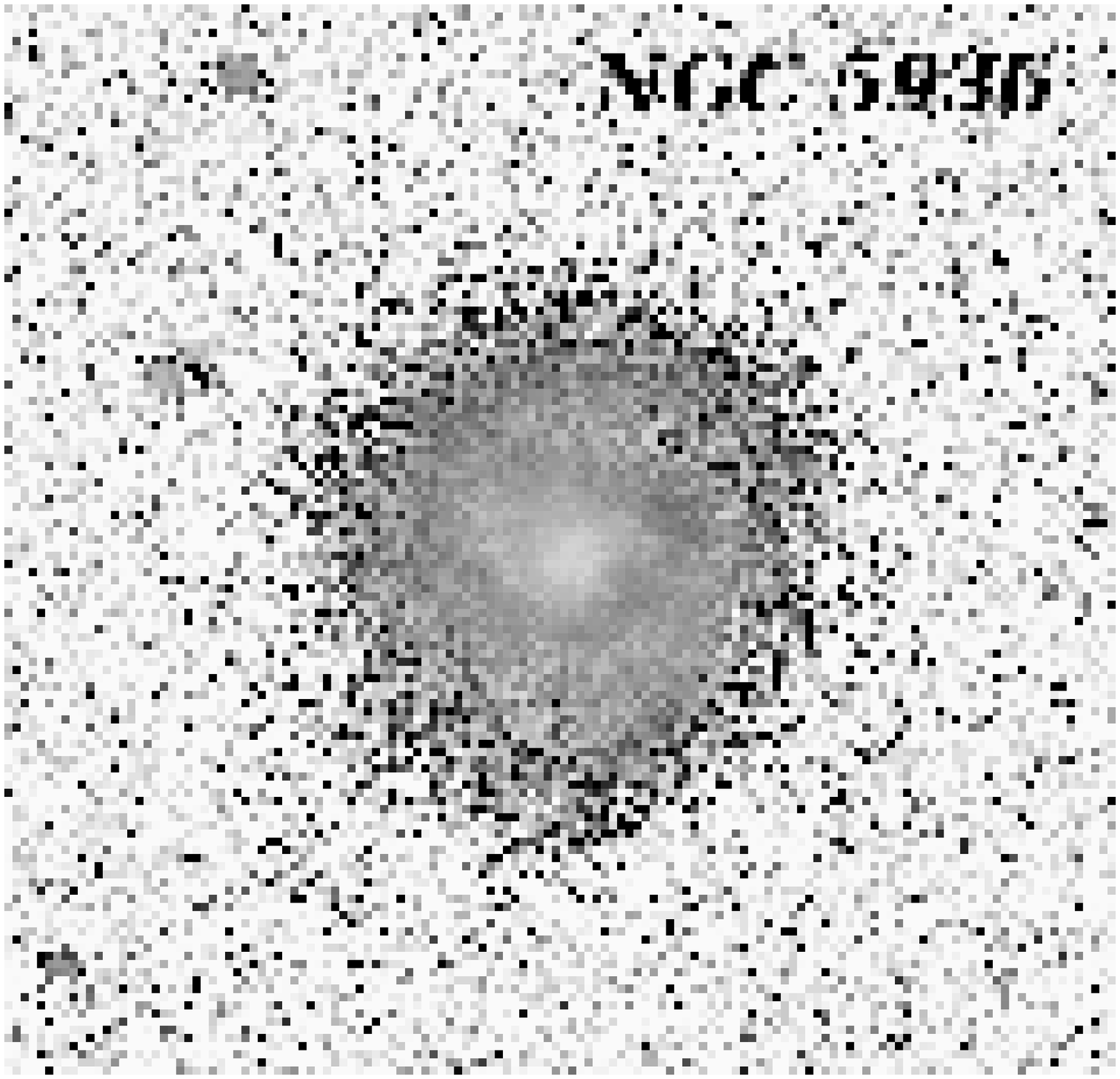}
\end{figure}

\newpage
\begin{figure}
\begin{center}
\includegraphics[width=10cm,keepaspectratio=true]{f5.eps}
\end{center}
\end{figure}

\newpage
\begin{figure}
\begin{center}
\includegraphics[width=10cm,keepaspectratio=true]{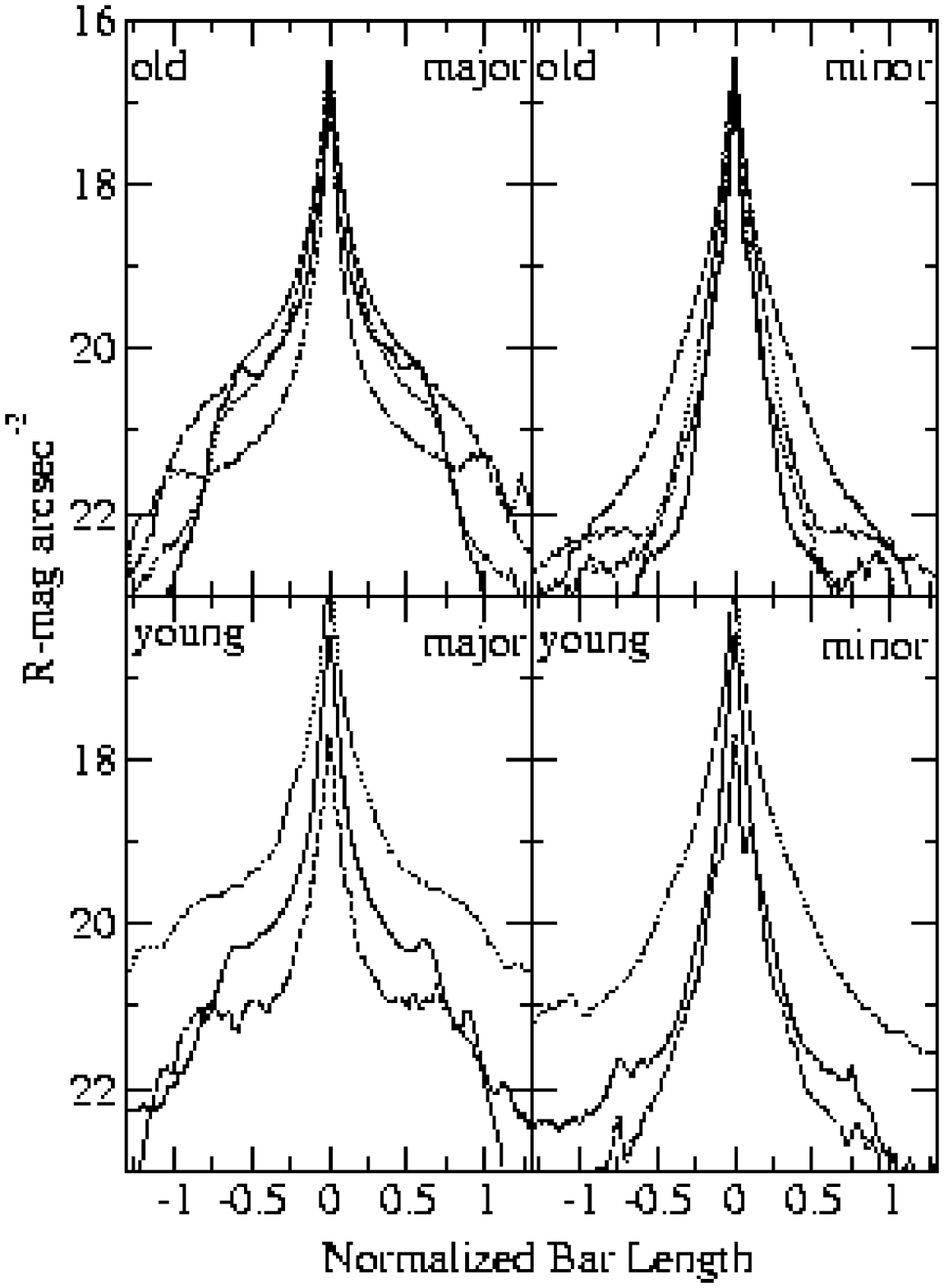}
\end{center}
\end{figure}


\begin{thebibliography}{}
\bibitem[Athanassoula(2002)]{ath02b} Athanassoula, E. 2002, \apjl, 569, L83
\bibitem[Athanassoula(2003)]{ath03} Athanassoula, E. 2003, \mnras, 341, 1179
\bibitem[Athanassoula(2005)]{ath05} Athanassoula, E. 2005, \mnras, 358, 1477
\bibitem[Athanassoula \& Martinet(1980)]{ath80} Athanassoula, E., \& Martinet, L. 1980, \aap, 87, L10
\bibitem[Athanassoula \& Misiriotis(2002)]{ath02a} Athanassoula, E., \& Misiriotis, A. 2002, \mnras, 330, 35
\bibitem[Berentzen et al.(2004)]{ber04} Berentzen, I., Athanassoula, E., Heller, C. H., \& Fricke, K. J.
2004, \mnras, 347, 220
\bibitem[Bournaud \& Combes(2002)]{bou02} Bournaud, F., \& Combes, F. 2002, \aap, 392, 83
\bibitem[de Jong(1996b)]{dej96b} de Jong, R. S. 1996, \aap, 313, 377
\bibitem[de Vaucouleurs et al.(1991)]{dev91} de Vaucouleurs, G., de Vaucouleurs, A., Corwin, H. G.,
Buta, R. J., Paturel, G., \& Fouque, P. 1991, Third Reference Catalog of Bright Galaxies (New York:
Springer-Verlag {\bf (RC3)}
\bibitem[Eggen, Lynden-Bell \& Sandage(1962)]{egg62} Eggen, O. J., Lynden-Bell, D., \& Sandage,
A. R. 1962, \apj, 136, 748
\bibitem[Elias et al.(1982)]{eli82} Elias, J. H., Frogel, J. A., Matthews, K., \& Neugebauer, G. 1982, \aj,
87, 1029
\bibitem[Elmegreen(1998)]{elm98} Elmegreen, D. M. 1998, Galaxies and Galactic Structure
(Englewood Cliffs: Prentice Hall)
\bibitem[Elmegreen \& Elmegreen(1985)]{elm85} Elmegreen, B. G., \& Elmegreen, D. M. 1985, \apj, 288, 438
\bibitem[Elmegreen, Elmegreen \& Hirst(2004)]{elm04} Elmegreen, B. G., Elmegreen, D. M., \& Hirst, A. C.
2004, \apj, 612, 191
\bibitem[Erwin(2004)]{erw04} Erwin, P. 2004, \aap, 415, 941
\bibitem[Erwin(2005)]{erw05} Erwin, P. 2005, \mnras, in press, (astro-ph/0508590)
\bibitem[Erwin \& Sparke(2002)]{erw02} Erwin, P., \& Sparke, L. S. 2002, \aj, 124, 65
\bibitem[Erwin \& Sparke(2003)]{erw03} Erwin, P., \& Sparke, L. S. 2003, \apjs, 146, 299
\bibitem[Friedli \& Benz(1995)]{fri95} Friedli, D., \& Benz, W. 1995, \aap, 301, 649
\bibitem[Fukugita, Shimasaku \& Ichikawa(1995)]{fuk95} Fukugita, M., Shimasaku, K., \& Ichikawa, T.
1995, \pasp, 107, 945
\bibitem[Gadotti \& de Souza(2003a)]{gad03a} Gadotti, D. A., \& de Souza, R. E. 2003a, \apjl, 583, L75
\bibitem[Gadotti \& de Souza(2003b)]{gad03b} Gadotti, D. A., \& de Souza, R. E. 2003b, \apss, 284, 527
\bibitem[Gadotti \& de Souza(2005)]{gad04a} Gadotti, D. A., \& de Souza, R. E. 2005, \apj, 629, 797
{\bf (Paper I)}
\bibitem[Gadotti \& de Souza(2004)]{gad04b} Gadotti, D. A., \& de Souza, R. E. 2004, in
The Interplay among Black Holes, Stars and ISM in Galactic Nuclei, Th. Storchi Bergmann, L. C. Ho,
H. R. Schmitt, eds., IAU Symp. 222, 423
\bibitem[Gadotti \& dos Anjos(2001)]{gad01} Gadotti, D. A., \& dos Anjos, S. 2001, \aj, 122, 1298
\bibitem[Gerssen, Kuijken \& Merrifield(2003)]{ger03} Gerssen, J., Kuijken, K., \& Merrifield, M. R.
2003, \mnras, 345, 261
\bibitem[Giovanelli et al.(1994)]{gio94} Giovanelli, R., Haynes, M. P., Salzer, J. J., Wegner, G., da
Costa, L. N., \& Freudling, W. 1994, \aj, 107, 2036
\bibitem[Jogee(2004)]{jog04a} Jogee, S. 2004, LNP Volume on AGN Physics on All Scales, Chapter 6, in press
(astro-ph/0408383)
\bibitem[Jogee et al.(2004)]{jog04b} Jogee, S. et al. 2004, \apjl, 615, L105
\bibitem[Jungwiert, Combes \& Axon(1997)]{jun97} Jungwiert, B., Combes, F., \& Axon, D. J. 1997, \aaps, 125, 479
\bibitem[Kitchin(1998)]{kit98} Kitchin, C. R. 1998, Astrophysical Techniques (Bristol: Inst. Physics)
\bibitem[Kormendy \& Kennicutt(2004)]{kor04} Kormendy, J., \& Kennicutt, R. C. 2004, \araa, 42, 603
\bibitem[Laine et al.(2002)]{lai02} Laine, S., Shlosman, I., Knapen, J. H., \& Peletier, R. F. 2002, \apj, 567, 97
\bibitem[Landolt(1983)]{lan83} Landolt, A. U. 1983, \aj, 88, 439
\bibitem[Laurikainen, Salo \& Buta(2004)]{lau04} Laurikainen, E., Salo, H., \& Buta, R. 2004, \apj, 607, 103
\bibitem[Maraston(1998)]{mar98} Maraston, C. 1998, \mnras, 300, 872
\bibitem[Martin(1995)]{mar95a} Martin, P. 1995, \aj, 109, 2428
\bibitem[Martin \& Roy(1995)]{mar95b} Martin P., \& Roy J.-R. 1995, \apj, 445, 161
\bibitem[Martin \& Friedli(1997)]{mar97} Martin, P., \& Friedli, D. 1997, \aap, 326, 449
\bibitem[Persson et al.(1998)]{per98} Persson, S. E., Murphy, D. C., Krzeminski, W., Roth, M., \& Rieke,
M. J. 1998, \aj, 116, 2475
\bibitem[Phillips(1993)]{phi93} Phillips, A. C. 1993, Ph.D. thesis, Univ. Washington
\bibitem[Phillips(1996)]{phi96} Phillips, A. C. 1996, in ASP Conf. Ser. 91, Barred Galaxies,
ed. R. Buta, D. A. Crocker, \& B. G. Elmegreen (IAU Colloq. 157; San Francisco: ASP), 44
\bibitem[Roberts \& Haynes(1994)]{rob94} Roberts, M. S., \& Haynes, M. P. 1994, \araa, 32, 115
\bibitem[Schlegel, Finkbeiner \& Davis(1998)]{sch98} Schlegel, D. J., Finkbeiner, D. P., \& Davis, M. 1998,
\apj, 500, 525
\bibitem[Searle \& Zinn(1978)]{sea78} Searle, L., \& Zinn, R. 1978, \apj, 225, 357
\bibitem[Shen \& Sellwood(2004)]{she04} Shen, J., \& Sellwood, J. A. 2004, \apj, 604, 614
\bibitem[Spitzer \& Schwarzschild(1951)]{spi51} Spitzer, L., \& Schwarzschild, M. 1951, \apj, 114, 385
\bibitem[Spitzer \& Schwarzschild(1953)]{spi53} Spitzer, L., \& Schwarzschild, M. 1953, \apj, 118, 106
\bibitem[Tinsley \& Gunn(1976)]{tin76} Tinsley, B. M., \& Gunn, J. E. 1976, \apj, 203, 52
\bibitem[Wozniak et al.(1995)]{woz95} Wozniak, H., Friedli, D., Martinet, L., Martin, P., \& Bratschi, P. 1995,
\aaps, 111, 115
\end{thebibliography}
\end{document}